\documentclass[usenatbib,preprint]{mn2e}

\usepackage[pdftex]{graphicx}
\usepackage {natbib,aas_macros}
\usepackage {mediabb}
\usepackage {bm}
\usepackage {amsmath,amssymb}
\usepackage {setspace}

\newcommand{\msun}{\thinspace M_\odot} 
\newcommand{\gcm}{~{\rm g~cm}^{-3} } 

\newcommand{\rR}{\frac{r}{R_0}}

\newcommand{\magB}{\mathbf{B}}

\newcommand{\vel}{\mathbf{v}}
\newcommand{\rad}{\mathbf{r}}

\newcommand{\ms}{~{\rm m} ~{\rm s}^{-1} } 
\newcommand{\kms}{~{\rm km} ~{\rm s}^{-1} }

\newcommand{\mum}{{\rm \mu} {\rm m} }

\newcommand{\msunyear}{\thinspace M_\odot~{\rm yr}^{-1}} 
\title{Early evolution of disk, outflow, and magnetic field of young stellar objects: Impact of dust model}

\author[Tsukamoto et al]{
Y. Tsukamoto$^{1}$,  M. N. Machida$^{2}$  H. Susa$^{3}$, H. Nomura$^{4}$, and  S. Inutsuka$^{5}$ \\
$^1$Graduate Schools of Science and Engineering, Kagoshima University, Kagoshima, Japan  \\
$^2$Department of Earth and Planetary Sciences, Kyushu University, Fukuoka, Japan \\
$^3$Department of Physics, Konan University, Okamoto, Kobe, Japan \\
$^4$Division of Science, National Astronomical Observatory of Japan, Mitaka, Tokyo, Japan \\
$^5$Department of Physics, Nagoya University, Aichi, Japan  \\
}

\begin{document}
\maketitle

\begin{abstract}
  The formation and early evolution of low mass young stellar objects (YSOs)
  are investigated using three-dimensional non-ideal magneto-hydrodynamics simulations.
  We investigate  the evolution of YSOs up to $\sim 10^4$ yr after protostar
  formation, at which protostellar mass reaches $\sim 0.1 \msun$.
  We particularly focus on the impact of the dust model on the evolution.
  We found that a circumstellar disk is formed in all simulations regardless of the dust model.
  Disk size is approximately 10 AU at the protostar formation epoch,
  and it increases to several tens of AU at $ \sim 10^4$ yr after protostar formation.
  Disk mass is comparable to central protostellar mass and gravitational instability develops.
  In the simulations with small dust size, the warp of the pseudodisk develops $\sim 10^4$ yr after protostar formation.
  The warp strengthens magnetic braking in the disk and decreases disk size.
  Ion-neutral drift can occur in the infalling envelope under the conditions that
  the typical dust size is $a \gtrsim 0.2 \mum$ and the protostar (plus disk) mass 
  is $M\gtrsim 0.1 \msun$.
  The outflow activity is anti-correlated to the dust size
  and the strong outflow appears with small dust grains.
\end{abstract}


\section{Introduction}
\label{intro}
Molecular cloud cores, which are birth places for protostars and
protoplanetary disks, are strongly magnetized.
Measurement of the Zeeman effect has shown that the
mass-to-flux ratio normalized
by its critical value is of order unity \citep{2008ApJ...680..457T,2012ARA&A..50...29C}.
The strong magnetic field of cloud cores is also supported
by three-dimensional simulations of molecular cloud formation \citep{2012ApJ...759...35I}.
This suggests that although the magnetic field is not strong enough
to support the core against gravitational collapse,
it should play a crucial role during a gravitational collapse of the core.
For example, angular momentum removal from the central region by the magnetic field
(so called magnetic braking) almost completely suppresses disk formation if
the neutral gas and magnetic field are well
coupled \citep[e.g.,][]{2008ApJ...681.1356M,2015MNRAS.452..278T}.

Another important feature of cloud cores is their low
ionization degree \citep[e.g.,][]{1990MNRAS.243..103U,2002ApJ...573..199N}.
The ionization degree of cloud cores ($\sim 10^{-20} \gcm$) is typically $ \sim 10^{-8}$,
and it decreases as density increases during the gravitational contraction phase.
In such a weakly ionized magnetized cloud, non-ideal effects, ohmic diffusion, the Hall effect,
and ambipolar diffusion play key roles.

It has been shown that ohmic diffusion largely affects the formation of young stellar objects.
Ohmic diffusion decouples the gas from the magnetic field
at $\rho>10^{-11} \gcm$ \citep{1990MNRAS.243..103U,1991ApJ...368..181N,2002ApJ...573..199N}
and gas can accrete to the central protostar leaving the magnetic flux at this density.
Furthermore, \citet{2011MNRAS.413.2767M} and \citet{2015MNRAS.452..278T} showed that
disk formation at the protostar formation epoch with size of  $r\sim 1$ to $10$ AU
is enabled by ohmic diffusion due to the decoupling.
However, note that the magnetic flux accumulates around the central dense region
of $\rho \lesssim 10^{-11} \gcm $ in the protostellar accretion phase
(the evolution phase after protostar formation), and
a huge amount of magnetic flux abides in the central compact region
when we neglect other non-ideal effects; this is because ohmic resistivity is an
increasing function of density and does not depend on the magnetic field strength.
Thus, neglecting other non-ideal effects may affect
evolution after protostar formation, during which
a large amount of magnetic flux is supplied towards the central region. 

Ambipolar diffusion, on the other hand,
plays a role not only in the high-density region but also in the low-density region of
$\rho \lesssim 10^{-11} \gcm$, and it becomes strong as magnetic flux accumulates.
Therefore it must play a central
role in magnetic field evolution in the protostellar evolution phase.
Previous studies have shown that ambipolar diffusion
further weakens the coupling between the magnetic field
and the gas in a newly born disk \citep{2015MNRAS.452..278T}.
Furthermore, magnetic flux can drift outwardly relative to the neutral motion 
in the envelope by ambipolar diffusion
\citep[e.g.,][]{2011ApJ...738..180L,2017PASJ...69...95T,2018MNRAS.473.4868Z}.
This outward magnetic field drift is a promising mechanism
to remove the magnetic flux from the central region in the protostellar accretion phase.
Furthermore, magnetic field drift possibly accompanies ion drift in the low-density
region, or more precisely, in the region where Hall
parameter is $\beta_{\rm Hall}\gg1$.
This may provide a unique observational opportunity to quantify
the magnetic field in the inner envelope of YSOs
\citep[see for recent attempt by ][]{2018A&A...615A..58Y}.

Because non-ideal effects (or finite conductivity) arise
due to the low ionization degree of the cloud core,
they inevitably depend on the microscopic properties of the gas.
Previous studies have shown that cosmic-ray ionization
and dust size distribution are the keys to determining the resistivity
of non-ideal effects \citep{1991ApJ...368..181N,2018MNRAS.478.2723Z,2019MNRAS.484.2119K}.
They are the source and sink of charged particles, respectively.

There are observations suggesting that dust growth may proceed even at
cloud core scale. For example, mid-infrared emission from the cloud core
is interpreted as scattered light of $\mum$ sized dust
grains \citep{2010A&A...511A...9S,2010Sci...329.1622P}.
In theoretical estimates of dust growth, dust size of $\lesssim 1 \mum$ 
is possibly realized within the free-fall time
in the dense inner part of the cloud core
\citep{2009A&A...502..845O,2013MNRAS.434L..70H}.
Thus, the impact of dust size on the non-ideal effect and cloud
core evolution should be investigated.

Recently, \citet{2016MNRAS.460.2050Z,2018MNRAS.473.4868Z} suggested
that dust growth and removal of the small dust grains
from MRN dust size  distribution ("truncated MRN") changes
magnetic resistivity significantly and enables disk formation.
On the other hand, several studies have reported disk formation with
small sized dust grains \citep{2016A&A...587A..32M,2016MNRAS.457.1037W,
  2015ApJ...801..117T,2017PASJ...69...95T,2018MNRAS.480.4434W, 2019MNRAS.486.2587W}.
Therefore, there are inconsistencies among previous studies.
\citet{2018MNRAS.473.4868Z} employed several simplifications for
numerical treatments to avoid small time stepping such as 
setting upper limit on ambipolar resistivity and
Alfv\`en velocity, neglecting Ohmic diffusion, and
imposing relatively large inner boundary of $2$ AU.
Note that large sink radius of $\gtrsim 1$ AU
tends to suppress disk formation \citep{2014MNRAS.438.2278M}.
We speculate that these treatments may have
some impacts on disk formation and early evolution.

Thus, we believe that an additional study investigating the
impact of dust models is required.
In this study, we investigated the early disk evolution phase
up to $\sim 10^4$ yr after protostar formation
particularly focusing on the impact of dust size difference.

\section{Numerical Method and Initial Conditions}
\label{method}
\subsection{Numerical Method}
In our numerical simulations, non-ideal magneto-hydrodynamics (MHD) equations were solved.
\begin{eqnarray}
\frac{D \mathbf{v}}{D t}&=&-\frac{1}{\rho}\left\{ \nabla
\left( P+\frac{1}{2}|\magB|^2 \right) - \nabla \cdot (\mathbf{ B B})\right\} \nonumber \\
- \nabla \Phi,  \\
\frac{D {\mathbf B}}{D t}  &=& \left( {\mathbf B} \cdot \nabla \mathbf{v} -  {\mathbf B} (\nabla \cdot \mathbf{v})\right)    \nonumber \\ 
&-& \nabla \times \left\{ \eta_O (\nabla \times \mathbf B)  \right.    \nonumber \\
&-& \left. \eta_A ((\nabla \times \mathbf B) \times \mathbf {\hat {B}}) \times \mathbf {\hat {B}}\right\},  \\
\nabla^2 \Phi&=&4 \pi G \rho,\\
\label{EOS}
P &=& P(\rho)= c_{\rm s, iso}^2 \rho \left\{ 1+\left(\frac{\rho}{\rho_{\rm crit}}\right)^{2/3} \right\}.
\end{eqnarray}
where $\rho$ is  gas density, 
$P$ is gas pressure,  
$\magB$ is the magnetic field,
${\mathbf {\hat {B}}}$ is defined as ${\mathbf {\hat {B}}}\equiv \magB/|\magB| $.
$\eta_O$ and $\eta_A$ 
are the resistivity for ohmic and ambipolar diffusion, respectively,
and $\Phi$ is the gravitational potential.
$G$ is a gravitational constant.
We adopted a barotropic equation of state (EOS)
in which gas pressure only depends on density.
$c_{\rm s, iso}=190 \ms$ is isothermal sound velocity at $10$ K.
We used a  critical density
of $\rho_{\rm crit}=4\times 10^{-14} \gcm$ above which gas behaves adiabatically.
In this study, we ignored the Hall effect.

We used the smoothed particle magneto-hydrodynamics (SPMHD) method to solve the equations
\citep{2011MNRAS.418.1668I,2013ASPC..474..239I}.
Our numerical code was parallelized with Message Passing Interface (MPI).
We treated ohmic and ambipolar diffusion according to the prescription described in \citet{2013MNRAS.434.2593T}.
Both the diffusion processes were  accelerated by super-time stepping (STS) \citep{Alexiades96}.

To calculate the time evolution after protostar formation,
we employed the sink particle technique \citep{1995MNRAS.277..362B}.
The sink particle was dynamically introduced when the density exceeds
$\rho_{\rm sink}=4 \times 10^{-13} \gcm$.
In our simulations, one sink particle was permitted.
The sink particle absorbs SPH particles
with $\rho>\rho_{\rm sink}$ within $r<r_{\rm sink}=1$ AU, and 
the mass and linear momentum of SPH particles are added to those of sink particle.
The sink particle interacts with SPH particles via gravity. 
The system was integrated up to
$\sim 10^{4}$ yr after protostar formation, at which
the mass of the sink particle (or protostar) reached $\sim 0.1 \msun$.

\subsection{Resistivity model}
We used the tabulated resistivity calculated by the methods described in \citet{2015ApJ...801...13S}.
We considered the ion species of 
${\rm H_3^+,H_2^+,H_3^+,HCO^+,Mg^+}$ 
${\rm  He^+,C^+,O^+,O_2^+,H_3O^+,OH^+,H_2O^+}$ and the neutral species of 
${\rm H,H_2, He, CO, O_2, Mg, O, C, HCO, H_2O, OH}$.
We also considered the neutral and singly charged dust grains, $g^0,~g^-,~g^+$.
We took into account cosmic-ray ionization,  gas-phase and dust-surface recombination, and ion-neutral reactions.
We also considered the indirect ionization by high-energy photons emitted by direct cosmic-ray ionization
(described as CRPHOT in the UMIST database).
The initial abundance and reaction rates were taken from the UMIST2012 database \citep{2013A&A...550A..36M}.
The grain-ion and grain-grain collision rates 
were calculated using the equations of \citet{1987ApJ...320..803D}.
The chemical reaction network was solved using the CVODE package \citep{hindmarsh2005sundials}.
We assumed that the system was in chemical equilibrium, which is valid
in the nearby star forming cloud as discussed in \citet{2016A&A...592A..18M}.
We calculated resistivity using the abundances of charged species in the equilibrium state.
The momentum transfer rate between neutral and charged species was
calculated using the equations described in \citet{2008A&A...484...17P}.
The temperature for the chemical network was
assumed to be $T=10(1+\gamma_T (\rho/\rho_c)^{(\gamma_T-1)}) ~{\rm K}$, where $\gamma_T=7/5$.
The dust internal density is fixed to be $\rho_d=2 \gcm$.
The cosmic ray ionization is fixed to be $\xi_{\rm CR}=10^{-17} {\rm s^{-1}}$.

For the dust models, we considered single sized dust models
and models with a size distribution of $n(a)\propto a^{-3.5}$.
For the single sized dust models, we considered
three dust size of of $a=0.035 \mum$, $0.1 \mum$, and $0.3 \mum$.
For the models with size distributions,
we considered models with minimum and maximum dust sizes of $a_{\rm min}=0.005 \mum$ and $a_{\rm max}=0.25 \mum$ (MRN distribution) 
and $a_{\rm min}=0.1 \mum$ and $a_{\rm max}=0.25 \mum$ \citep[truncated MRN distribution
which mimics the size distribution of][]{2018MNRAS.473.4868Z}.
We assumed a fixed dust-to-gas mass ratio of 0.01.

\subsection{Initial conditions}
\label{init_condition}
We adopted the density-enhanced Bonnor-Ebert sphere surrounded by medium with a steep density
profile of $\rho \propto r^{-4}$ as the initial density profile,
\begin{eqnarray}
  \rho(r)=\begin{cases}
  \rho_0 \xi_{\rm BE}(r/a) ~{\rm for} ~ r < R_c \\
  \rho_0 \xi_{\rm BE}(R_c/a)(\frac{r}{R_c})^{-4} ~{\rm for} ~ R_c < r < 10 R_c.
  \end{cases}
\end{eqnarray}
and
\begin{eqnarray}
  a=c_{\rm s, iso} \left( \frac{f}{4 \pi G \rho_0} \right)^{1/2}.
\end{eqnarray}
where $\xi_{\rm BE}$ is non-dimensional density profile of the
critical Bonnor-Ebert sphere,
$f$ is a numerical factor related to the strength of gravity,
and $R_c=6.45 a$ is the radius of the cloud core.
$f=1$ corresponds to the critical Bonnor-Ebert sphere, and
the core with $f>1$ is gravitationally unstable.

A Bonnor-Ebert sphere is determined by specifying central density $\rho_0$, the
ratio of the central density to density at $R_c$ $\rho_0/\rho(R_c)$, and $f$.
In this study, we adopted the values of $\rho_0=7.3\times 10^{-18} \gcm$,
$\rho_0/\rho(R_c)=14$, and $f=2.1$. Then, the radius of the core
is $R_c=4.8\times 10^3$ AU, and the enclosed mass within $R_c$ is $M_c=1 \msun$.
The $\alpha_{\rm therm}$ ($\equiv E_{\rm therm}/E_{\rm grav}$) is equal to $0.4$, where
$E_{\rm therm}$ and $E_{\rm grav}$ are the thermal and gravitational energy of the central core
(without surrounding medium), respectively.
The steep envelope was adopted to put the outer
boundary far away from the central cloud core.
With the steep profile, the total mass of the entire domain remains $\sim 2 M_c$.
For rotation of the cloud core, we adopted an angular velocity 
profile of $\Omega(d)=\frac{\Omega_0 }{\exp(10(d/(1.5 R_c))-1)+1}$
where $d=\sqrt{x^2+y^2}$ and  $\Omega_0=2.3\times 10^{-13} {\rm s^{-1}}$.
$\Omega(d)$ is almost constant for $d<1.5 R_c$ and rapidly decreases for $d > 1.5 R_c$.
The ratio of the rotational to gravitational energy $\beta_{\rm rot}$ within
the core is $\beta_{\rm rot}$ ($\equiv E_{\rm rot}/E_{\rm grav}$) $=0.03$,
where $E_{\rm rot}$ is the rotational energy of the core.

We constructed a magnetic field profile that is constant, 
has only the $z$ component at the center, and
asymptotically obeys $B\propto r^{-2}$ as $r \to \infty$
(see Appendix A for details of the magnetic field configuration).
The merit of this profile is the avoidance of a low $\beta$ region in the surrounding medium, which
appears when adopting constant magnetic field strength and radially decreasing density.
We assumed the characteristic length scale of the field configuration to be $R_0=R_c$.
The central magnetic field strength
and plasma $\beta$ were $B_0=62 \mu G$ and $\beta=1.6 \times 10^1$, respectively.
With our magnetic field profile, the mass-to-flux ratio of the core $\mu$
relative to the critical value was $\mu/\mu_{\rm crit}=(M_c/\Phi_{\rm mag})=8$,
where $\Phi_{\rm mag}$ is the magnetic flux of the core and $\mu_{\rm crit}=(0.53/3 \pi)(5/G)^{1/2}$.
The mass-to-flux ratio is large compared to the observed value because the magnetic field
becomes weak in the outer region with our magnetic field configuration. However, note that
if we assume a constant magnetic field profile with central magnetic
field strength which is widely used in
previous studies, the mass-to-flux ratio is $\mu_{\rm const}/\mu_{\rm crit}=4$.
This value would be more suitable
to compare our magnetic field strength with those of previous studies because
we investigated time evolution of the gas in the central region of the core.
We resolved 1 $\msun$ with $3\times 10^6$ SPH particles. Thus, each particle had a mass of
$m=3.3\times 10^{-7} \msun$.

The model names and corresponding dust sizes are summarized in Table 1.
Table 1 also summarizes simulation results.

\begin{table*}
\label{summary of result}
\begin{center}
  \caption{
    Model name, dust size and summary of the results are given.
    The model with a size distribution has the distribution of $n(a)\propto a^{-3.5}$.
    $ t_{\rm star} [{\rm yr}]$,
    $t_{\rm end} [{\rm yr}]$,
    $r_{\rm disk, end}$  [AU],
    $r_{\rm outflow, end}$  [AU], and 
    $r_{\rm drift, end}$  [AU]
    are the time at protostar formation,
    time at the end of the simulation,
    centrifugal radius at $t_{\rm end}$,
    outflow size  at $t_{\rm end}$, and 
    maximum radius at which magnetic field drift velocity is larger than $0.19 \kms$ at $t_{\rm end}$, respectively.
}		
\begin{tabular}{ccccccc}
\hline\hline
 Model name  & $a_{\rm dust} [\mum]$ &$ t_{\rm star} [{\rm yr}]$ & $t_{\rm end} [{\rm yr}]$ & $r_{\rm disk, end}$  [AU]& $r_{\rm outflow, end}$  [AU] & $r_{\rm drift, end}$ [AU] \\
\hline
model\_a0035$\mum$ & 0.035  & $4.15\times 10^4$ & $5.6\times 10^4$ &  29  & 5100 & $< 100$  \\
model\_a01$\mum$ & 0.1  &  $4.15\times 10^4$ &$5.3\times 10^4$&  18  &      2300 & $< 100$   \\
model\_a03$\mum$ & 0.3  & $4.15\times 10^4$ &$5.6\times 10^4$&  59  &       3700 & 780   \\
model\_trMRN  & $0.1<a<0.25$ & $4.15\times 10^4$  &$5.3\times 10^4$& 53  &  3100 &  770   \\
model\_MRN  & $0.005<a<0.25 $ & $4.15\times 10^4$ & $5.3\times 10^4$ & 23   &  2400 & $< 100$ \\
\hline
\end{tabular}
\end{center}
\footnotesize
\end{table*}

\section{Results}
\subsection{Impact of dust size on resistivity}
\label{resistivity}
\begin{figure*}
  \includegraphics[width=50mm,angle=-90]{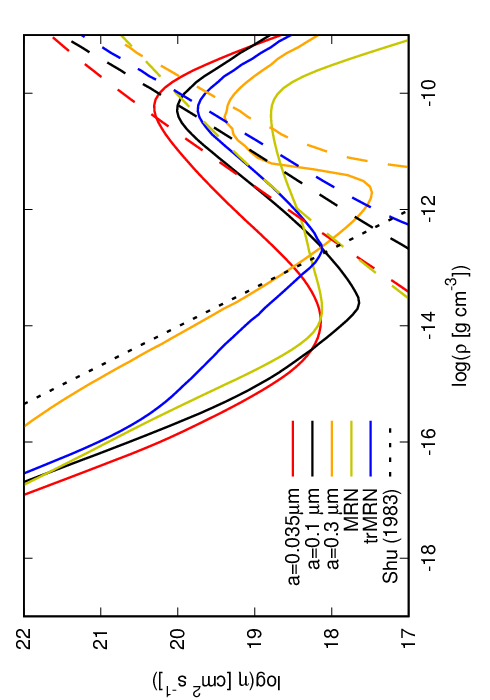}
  \includegraphics[width=50mm,angle=-90]{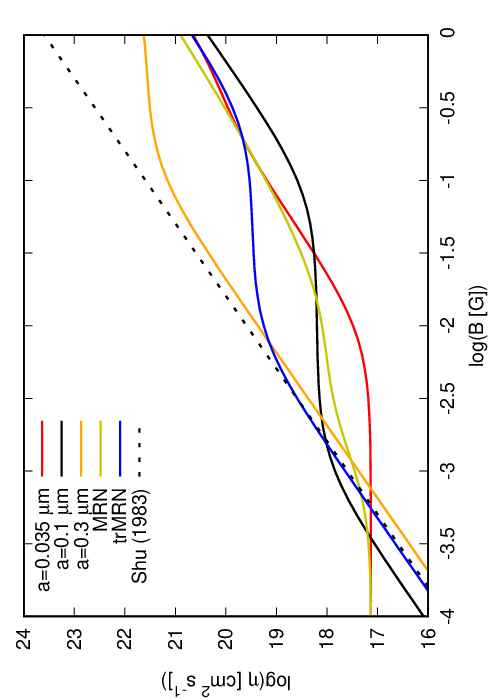}
\caption{
  The left panel shows $\eta_A$ (solid lines) and $\eta_O$ (dashed lines)
  as a function of density with a fixed magnetic field strength of  $B=30 {\rm mG}$.
  The right panel shows $\eta_A$ as a function
  of the magnetic field with a fixed density of  $\rho=4\times 10^{-15} \gcm$.
  The dotted black line shows $\eta_A=B^2/(C\gamma \rho^{3/2})$ given by \citet{1983ApJ...273..202S}.
}
\label{eta_fig}
\end{figure*}

First, we investigate how resistivity depends on the dust model.
Figure \ref{eta_fig} shows $\eta_O$
and $\eta_A$ with different dust models.
The left panel of figure \ref{eta_fig} shows $\eta_O$
and $\eta_A$ as a function of density. This indicates
that an increase in dust size has two contradicting impacts on $\eta_A$.
With larger dust size, $\eta_A$  is large in the low-density
region ($\rho \lesssim10^{-13} \gcm$) and 
small in the high-density region ($\rho \gtrsim10^{-13} \gcm$).
Thus, dust growth has a positive impact on the
decoupling between magnetic field and gas in the
low-density region but has a negative impact in the high-density region.

The small  $\eta_A$ with the small dust grains in the low-density 
region is caused by an increase of ion abundance in the gas phase.
With small dust grains, the electrons in the gas phase are more efficiently absorbed 
by dust grains and ions lose their counterparts.
As a result, recombination rate in the gas phase decreases.
This causes an increase in ion abundance and decrease
of $\eta_A$ in the low-density region with small dust grains.
Note that, as dust size increases, $\eta_A$ becomes similar
to the analytic formula by \citet{1983ApJ...273..202S},
$\eta_A=B^2/(\gamma C \rho^{3/2})$
(dotted line) in the low-density region where
$\gamma=3.5 \times 10^{13} {\rm cm^3 (g~s)^{-1}}$ and $C=3\times 10^{-16} {\rm (g~cm^{-3})^{1/2}}$.
This is because the abundance and total surface area
of dust decrease as the dust size increases,
and the system becomes similar to that of dust-free case.
The mutually contradicting impact of dust
size on $\eta_A$ in the  high and low-density regions may
introduce diversity to the evolution of disk, outflow, and magnetic flux.

On the other hand,
$\eta_O$ monotonically decreases as dust size increases.
This is particularly clear from comparing
the monosize dust model.
This shows that ohmic diffusion becomes less important as dust grows in
the molecular cloud core or in the circumstellar disk.

The right panel of figure \ref{eta_fig} shows $\eta_A$ as a function
of the magnetic field at $\rho= 4 \times 10^{-15} \gcm$.
This shows that the dependence of $\eta_A$ on magnetic
field strength is not simple  particularly for small dust size.
$\eta_A$ of a=$0.035\mum$ is almost independent of the magnetic field in $B<10^{-2} {\rm G}$
(and hence behaves like "ohmic diffusion").
As the magnetic field increases, on the other hand, $\eta_A$ obeys $\eta_A\propto B^2$.
The change in dependence on magnetic field is related to the weak coupling between
charged particles and the magnetic field at this density and magnetic field
strength \citep[for more detail, see][]{2002ApJ...573..199N}.
As dust size increases, the plateau slides to higher magnetic
field strength (see blue and black lines) and becomes narrow.
The right panel clearly indicates that, although
the relation of $\eta_A \propto B^2$ is often assumed for ambipolar diffusion,
this is not always valid.

\subsection{Time evolution of fiducial models}
\label{fiducial_model}
In this section, we describe the time evolution of two fiducial models:
one with small dust grains and one with large dust grains.
Important features, which will be discussed in subsequent sections, will be highlighted here.

\subsubsection{Time evolution of a model with small dust grains}
\begin{figure*}
\includegraphics[width=110mm,,angle=-90]{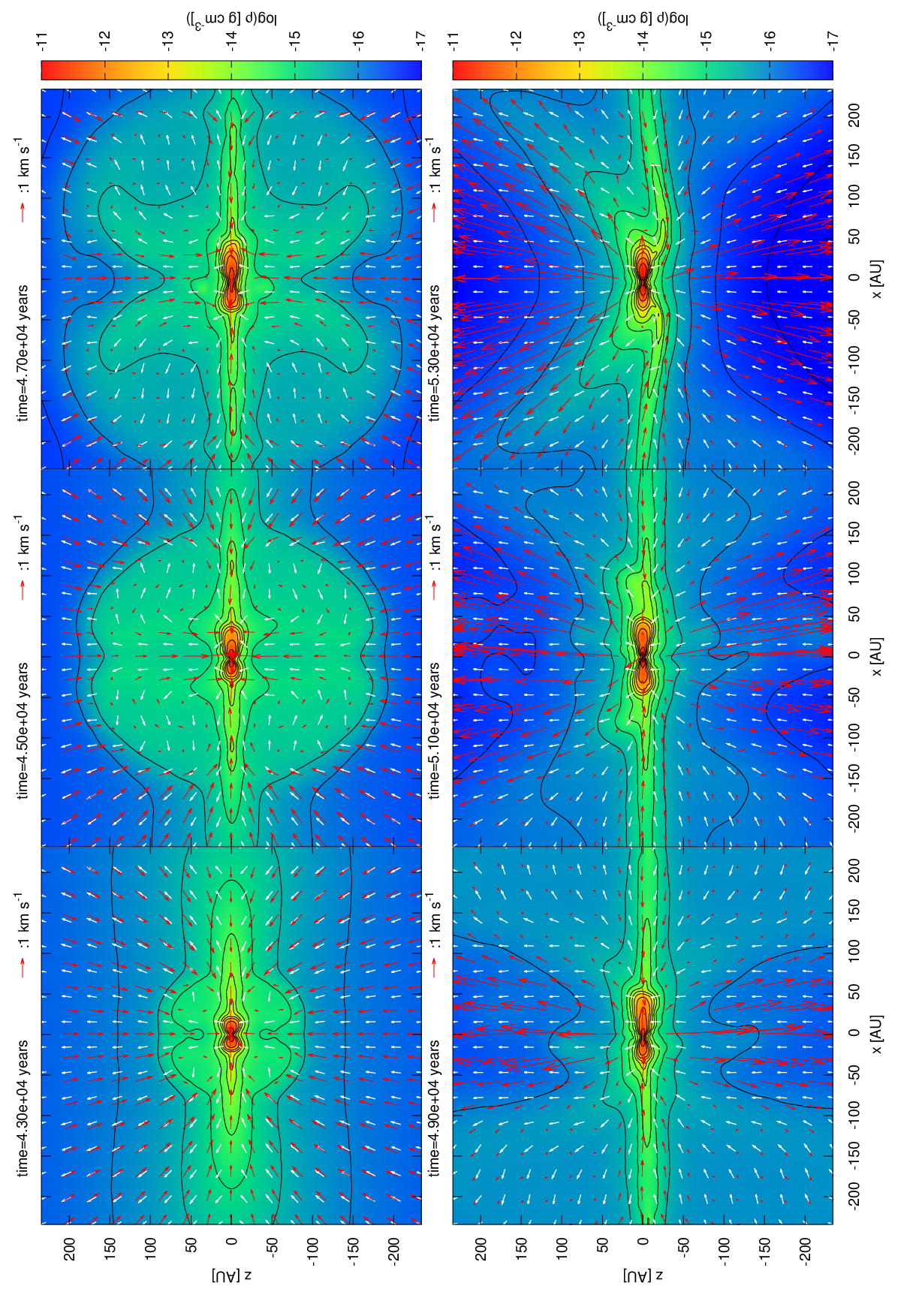}
\caption{
  Density cross-sections on the $x$-$z$ plane for central 500-AU
  square region of model\_MRN at $t=4.3 \times 10^4$, $4.5 \times 10^4$, $4.7 \times 10^4$,$4.9 \times 10^4$, $5.1 \times 10^4$, and $5.3 \times 10^4$ yr.
  A protostar is formed at $t=4.1\times 10^4$ yr.
  Red arrows show the velocity field, and white arrows show the direction of the magnetic field.
}
\label{density_xz_small_MRN}
\end{figure*}

\begin{figure*}
\includegraphics[width=110mm,,angle=-90]{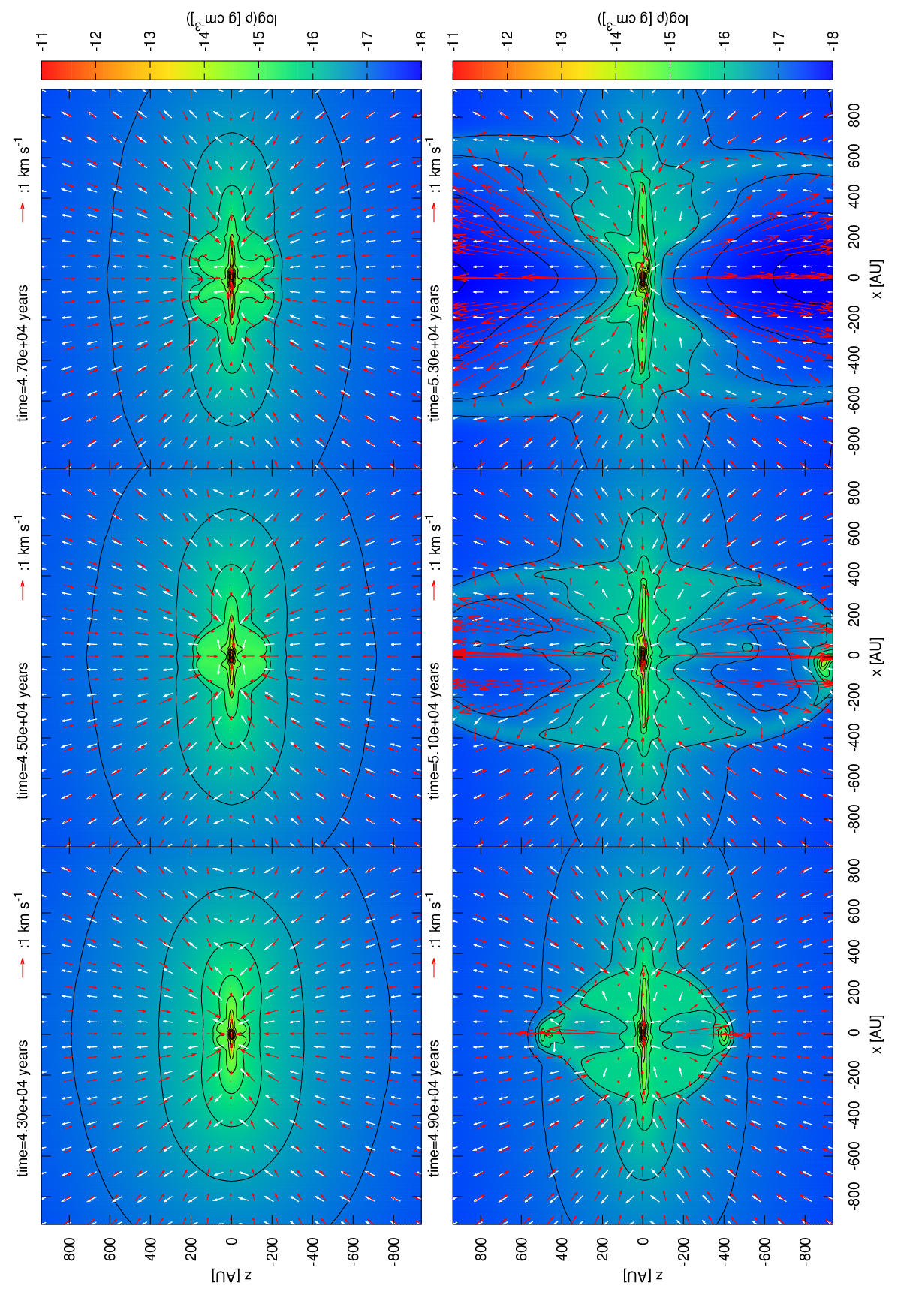}
\caption{
  Same as figure \ref{density_xz_small_MRN} but for central 2000 AU square region of model\_MRN.
}
\label{density_xz_large_MRN}
\end{figure*}

First, we investigate model\_MRN as a fiducial model with small dust grains.
Figures \ref{density_xz_small_MRN} and \ref{density_xz_large_MRN}
show the density evolution in the 500 AU scale box and
in the 2000 AU box of model\_MRN, respectively.
In this model, the protostar is formed at $t=4.1\times 10^4$ yr.
The top left panel of figure \ref{density_xz_small_MRN} shows that
a weak outflow with $v< 1 \kms$ is formed.
This outflow expands to $150$ AU in size at $t=4.5\times 10^4$ yr (top middle)
but almost stalls (top left).
Then a stronger and more collimated outflow with $v>1 \kms$ is launched (bottom left).
At $t=5.1 \times 10^{4}$ yr (which corresponds to $\sim 10^4$ yr after protostar formation),
fast outflow dominates in the envelope (bottom middle).

  The slow outflow is driven by the first core and the fast outflow is driven by the circumstellar disk.
  In our simulations, the stellar outflow is not resolved due to the sink particle with radius of $1$ AU.
  The outflow velocity is roughly determined by rotation velocity  (i.e., Keplerian velocity;
  $v_{\rm K} \sim 1 \kms (100/{\rm AU})^{-1/2}(M_{\rm star}/0.1 M_\odot)^{1/2}$ ) at the launching point.
  This suggests that the fast outflow is launched at several tens of AU corresponding to the disk radius.

The bottom middle panel of figure \ref{density_xz_large_MRN}
shows that the outflow head reaches $\sim 1000$ AU at this epoch.
The property of outflow will be investigated in more detail in \S \ref{outflow_evolution}.
We find that the warp of the pseudo-disk develops at $r\sim 100$ AU,
approximately  $10^4$ yr after protostar formation (bottom right panel).
This pseudo-disk warp strengthens the magnetic field in the disk and negatively impacts disk growth.
We will revisit pseudo-disk warp  in more detail in \S \ref{warp}.

\begin{figure*}
\includegraphics[clip,trim=0mm 0mm 0mm 0mm,width=110mm,,angle=-90]{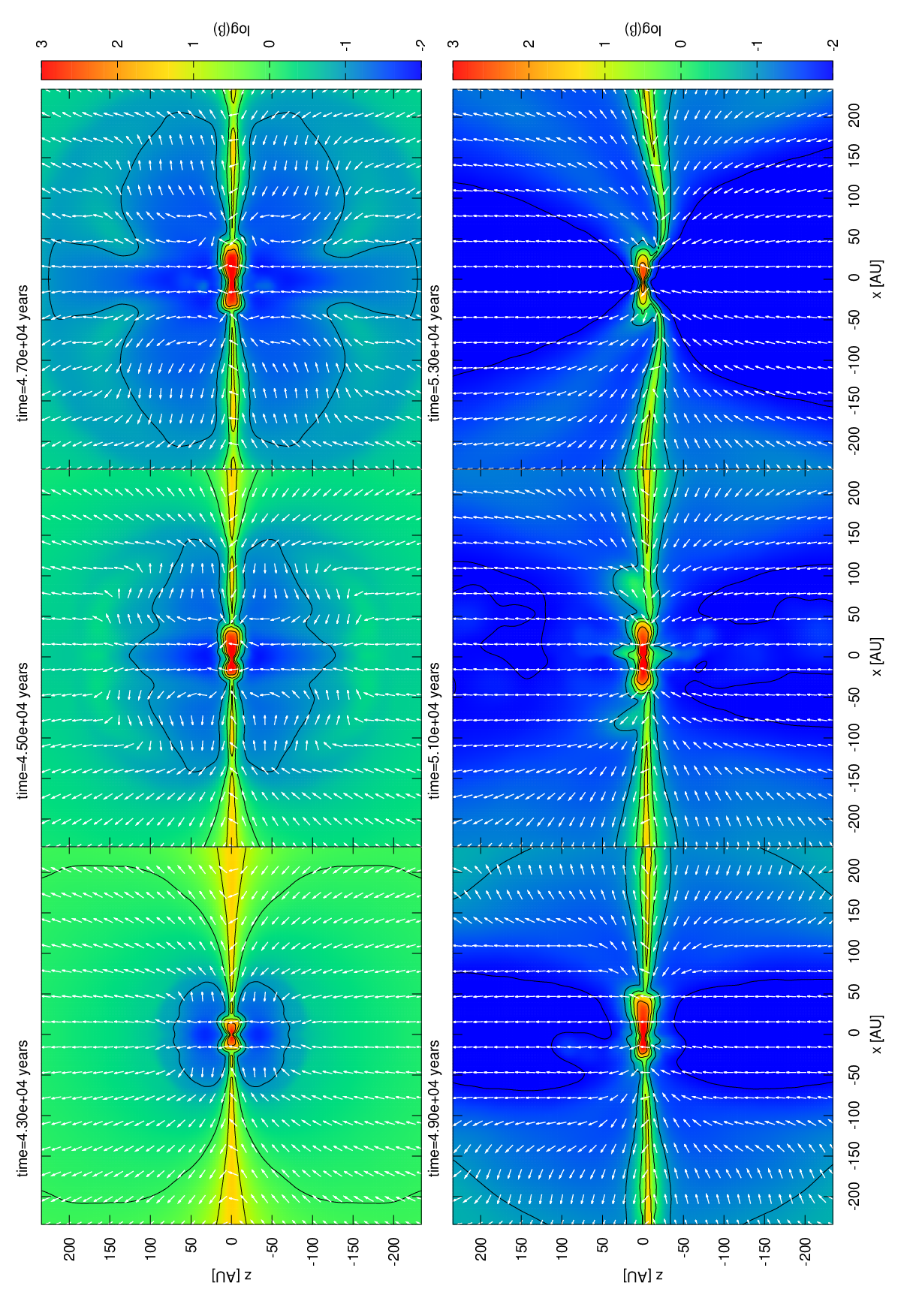}
\caption{
  Same as figure \ref{density_xz_small_MRN} but with cross-sections of plasma $\beta$ of model\_MRN.
}
\label{beta_xz_small_MRN}
\end{figure*}

Figure \ref{beta_xz_small_MRN} shows the plasma $\beta$ map on the $x$-$z$ plane in the  500 AU scale box.
The plasma $\beta$ has a large dynamic range from $\beta > 10^3$ in
the disk to $\beta < 10^{-2}$ in the upper envelope.
At the protostar formation epoch (top left panel),
the low $\beta$ region is localized in $r<100$ AU, and
it expands as time proceeds.

The white arrows around the midplane
(around the $x$ axis) indicate that the magnetic field is highly pinched
towards the center, and  the so-called "hour-glass" magnetic field configuration is realized.
The white arrows in the bottom right panel show that the "neck" of
the hour-glass magnetic field configuration shifts towards $z<0$
and contracts as the warp develops.
The warp of the pseudo-disk is more clearly seen in this $\beta$ map.
This contraction enhances the magnetic field in the disk. As a result,
the plasma $\beta$ of the disk decreases from $\beta \gtrsim 10^3$ to $\beta\sim 10^2$
once the warped pseudo-disk develops (from bottom middle to bottom right panels).

\begin{figure*}
\includegraphics[clip,trim=0mm 0mm 0mm 0mm,width=110mm,angle=-90]{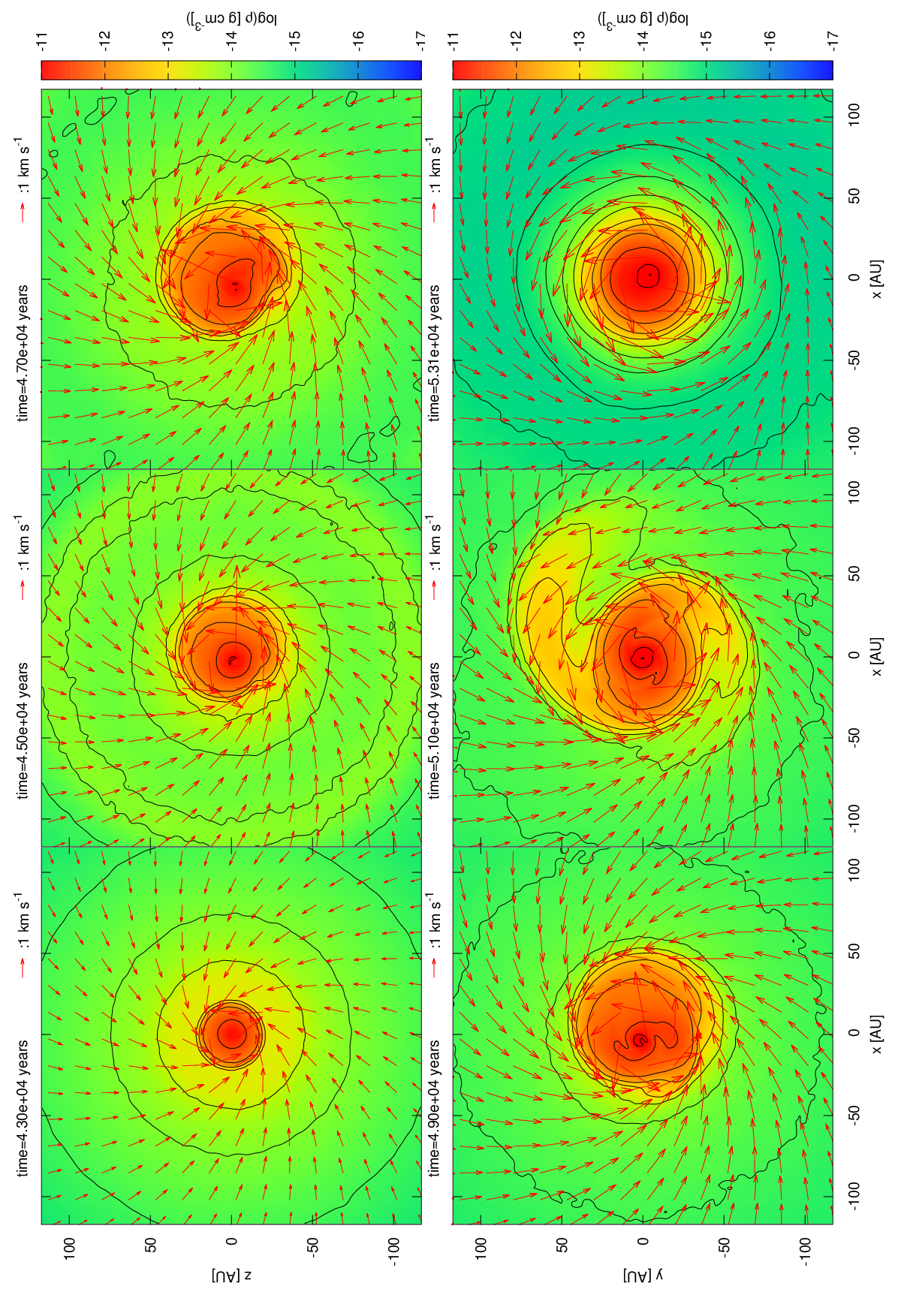}
\caption{
  Same as figure \ref{density_xz_small_MRN} but on the $x$-$y$ plane of model\_MRN.
}
\label{density_xy_small_MRN}
\end{figure*}

Figure \ref{density_xy_small_MRN} shows the density map on the $x$-$y$ plane
at the same epochs of figure \ref{density_xz_small_MRN} in the 250 AU scale box.
The central high-density region ($\rho \gtrsim 10^{-14}\gcm$) is the circumstellar disk.
In our simulations, the circumstellar disk forms immediately after protostar formation
(or sink particle creation) and survives for $t\sim 10^4$ yr thereafter.
Thus, our results are consistent with the disk formation scenario in
\citet{2011MNRAS.413.2767M} and \citet{2012PTEP.2012aA307I}
where the first core directly becomes the circumstellar disk.
The disk radius gradually grows from $\sim 10$ AU (top left) to $\sim 50$ AU size (bottom middle).
As the disk size increases, spiral arms develop due to gravitational instability.
We observed that the spiral arms are repeatedly formed in the disk.
The emergence of the spiral arms and outflow activity are
correlated, and strong outflow is launched when the spiral arms are prominent.
As the warp of the pseudo-disk develops, the disk begins to shrink (from bottom middle to right panels).
More rigorous analysis of disk size evolution is presented in \S \ref{disk_evolution}.

\subsubsection{Time evolution of a model with large dust grains}
\begin{figure*}
\includegraphics[clip,trim=0mm 0mm 0mm 0mm,width=110mm,angle=-90]{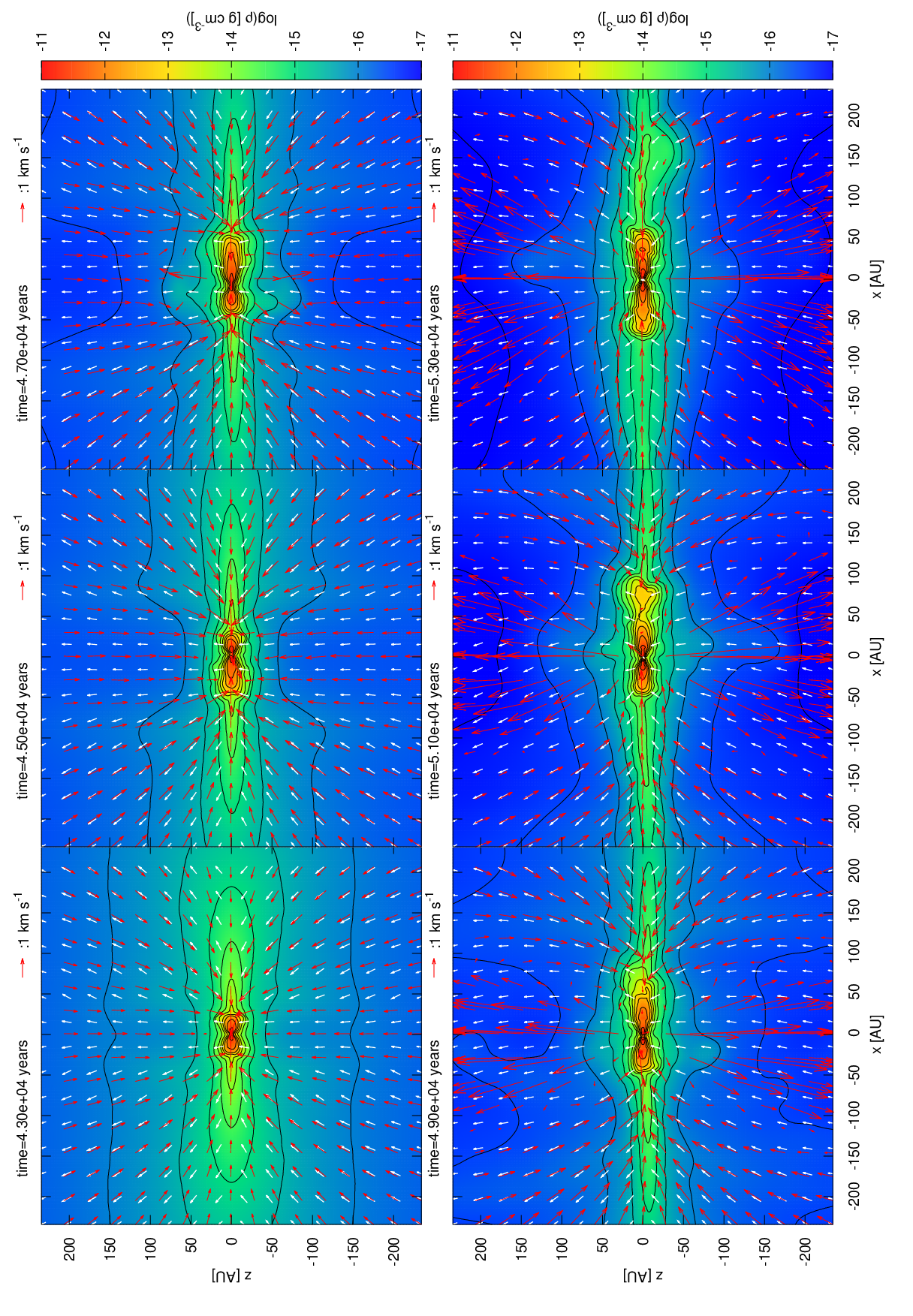}
\caption{
  Same as figure \ref{density_xz_small_MRN} but for model\_a03$\mum$.
}
\label{density_xz_small_a03mum}
\end{figure*}

\begin{figure*}
\includegraphics[clip,trim=0mm 0mm 0mm 0mm,width=110mm,angle=-90]{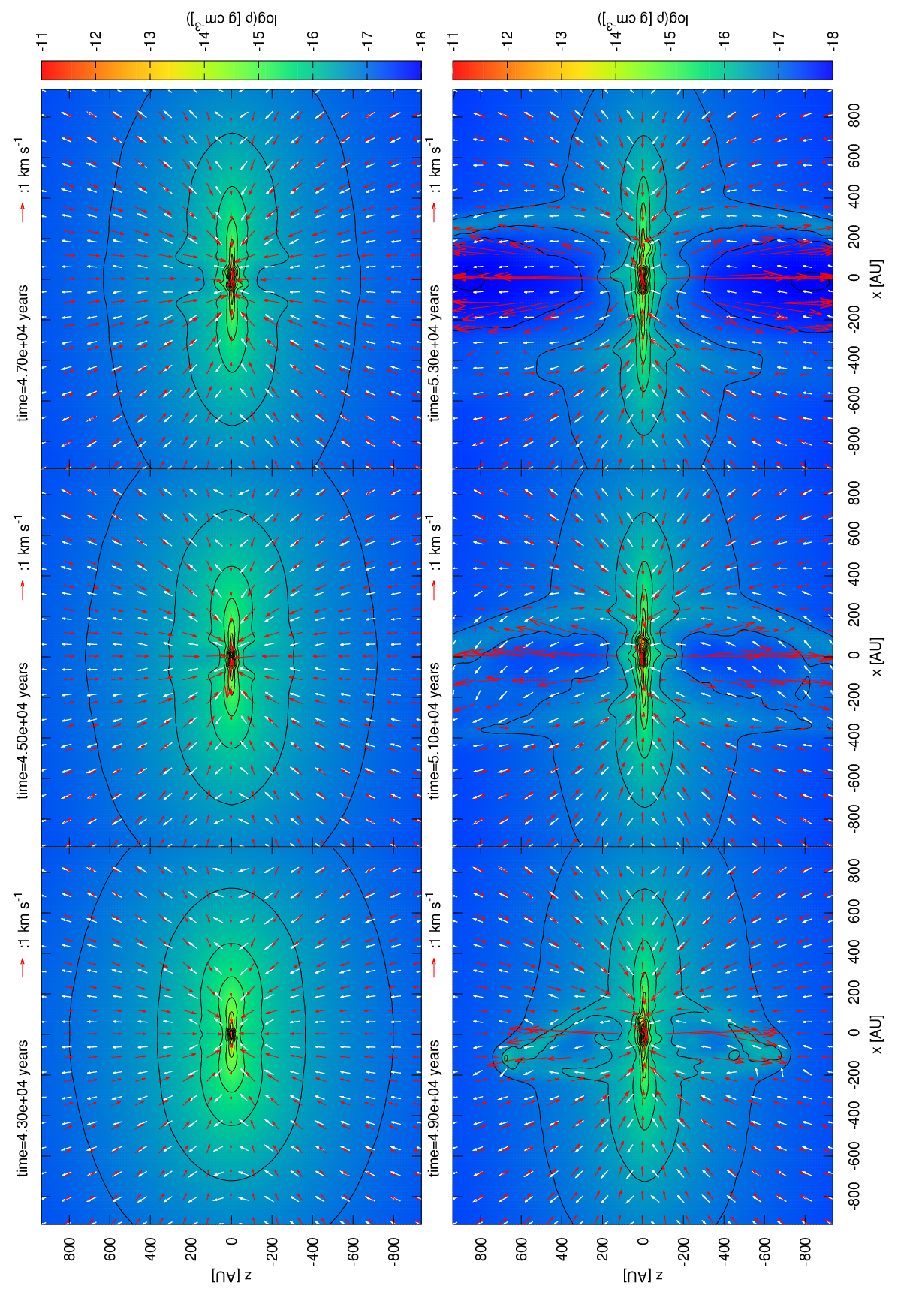}
\caption{
  Same as figure \ref{density_xz_large_MRN} but for model\_a03$\mum$.
}
\label{density_xz_large_a03mum}
\end{figure*}

Next, we will investigate the time evolution of a model with large dust grains,
model\_a03$\mum$.
Figures \ref{density_xz_small_a03mum} and \ref{density_xz_large_a03mum}
show the density evolution on the $x$-$z$
plane in the 500 AU and 2000 AU scale boxes of model\_a03$\mum$, respectively.
The clear difference between this model and model\_MRN at the protostar formation epoch
is the absence of outflow from first core (top left panels).
This is due to stronger ambipolar diffusion in the low-density region (see figure \ref{eta_fig}).
However, as time proceeds, the outflow with $v>1 \kms$ eventually
forms at $\sim 7\times 10^3$ yr after protostar formation (bottom left panels).
The bottom left panel of figure \ref{density_xz_large_a03mum} shows that the outflow is slightly tilted
from the $z$ direction. This is due to the non-axisymmetric spiral arms in the disk.
(see figure \ref{density_xy_small_a03mum}; non-axisymmetric m=1 mode frequently develops in this model).
In this model, the outflow is more collimated and its cylindrical radius ($\sqrt{x^2+y^2}$)
is smaller than that of model\_MRN at $t=5.3\times 10^4$ yr.
As shown \S \ref{outflow_evolution},
the outflow angular momentum of this model is much smaller than that of model\_MRN
and the smaller cylindrical radius is one reason for this difference.
Another important difference between these two models is the absence of pseudo-disk warp.
Even though we calculated the system evolution  $\sim 10^4$
yr after protostar formation, the
pseudo-disk warp (and any symptom of it) does not develop in this model.
Note that, in the case of model\_MRN, a weak warp is already
formed at $t=5.1 \times 10^4$ yr (bottom middle
panel of figure \ref{density_xz_small_MRN} and \ref{beta_xz_small_MRN}).
Due to the absence of pseudo-disk warp, the large disk
is maintained (see figure \ref{density_xy_small_a03mum}).

\begin{figure*}
\includegraphics[clip,trim=0mm 0mm 0mm 0mm,width=110mm,angle=-90]{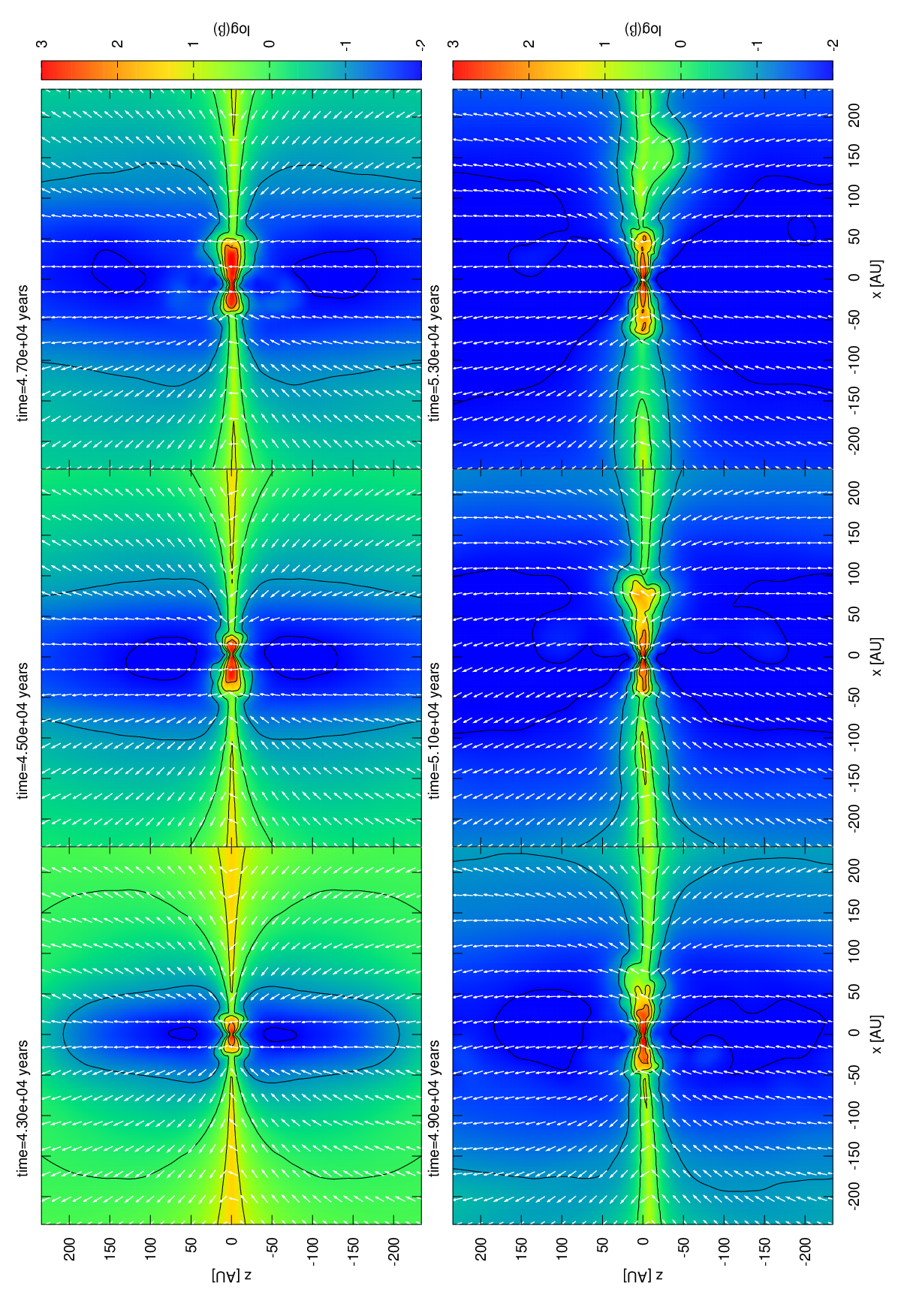}
\caption{
  Same as figure \ref{beta_xz_small_MRN} but for model\_a03$\mum$.
}
\label{beta_xz_small_a03mum}
\end{figure*}

\begin{figure*}
\includegraphics[clip,trim=0mm 0mm 0mm 0mm,width=110mm,angle=-90]{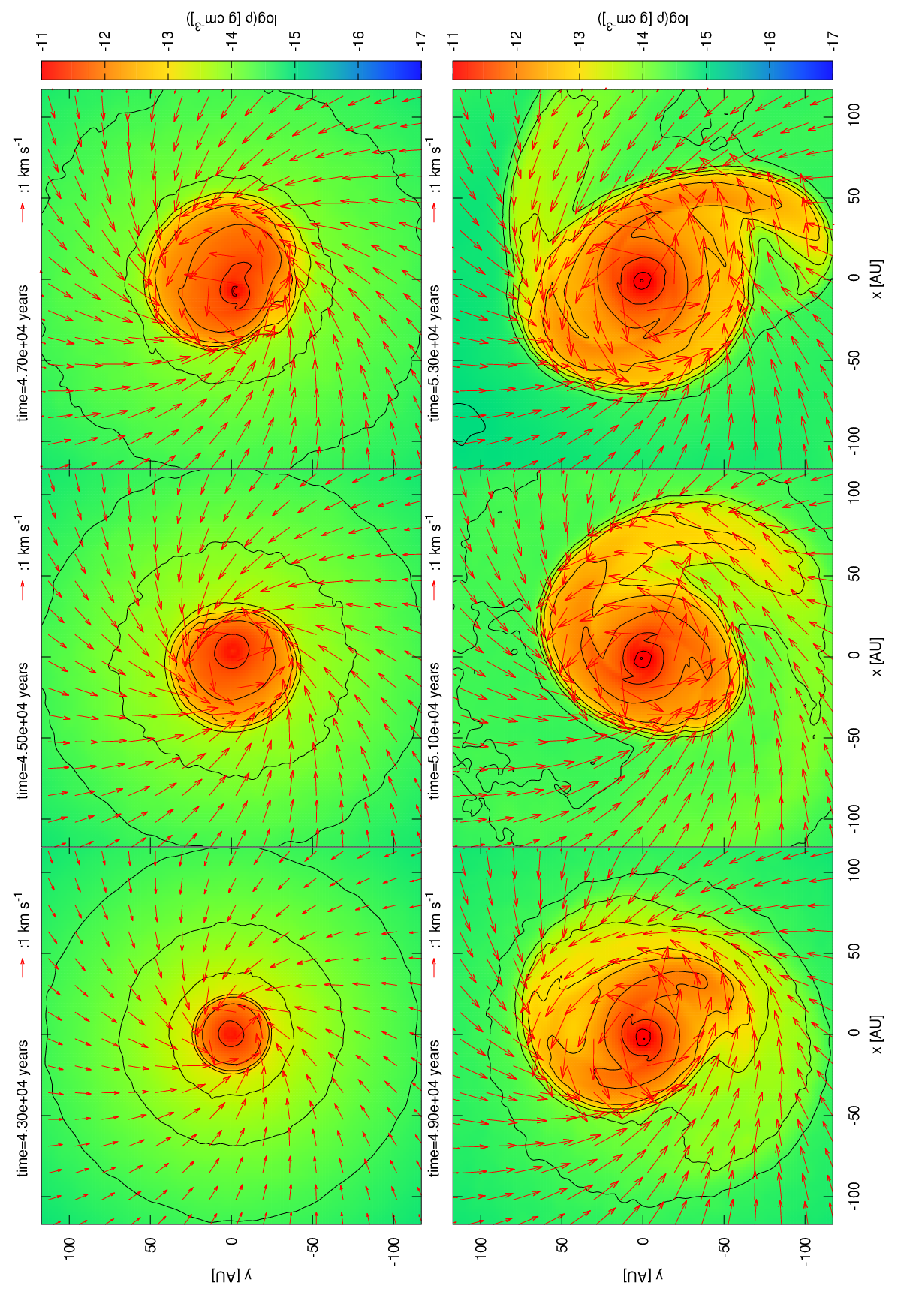}
\caption{
  Same as figure \ref{density_xy_small_MRN} but for model\_a03$\mum$.
}
\label{density_xy_small_a03mum}
\end{figure*}

Figure \ref{beta_xz_small_a03mum} shows the plasma $\beta$ map of model\_a03$\mum$.
The top left panel shows that a bipolar low $\beta$ structure already forms at the protostar
formation epoch,
but it does not drive the outflow until the magnetic field is sufficiently amplified by disk rotation.
An interesting difference between model\_MRN and model\_a03$\mum$
is the thickness of the current sheet at the midplane. The bottom left panel shows that the contours of
plasma $\beta$ at $r\gtrsim50$ AU are sparse around the midplane,
indicating that the magnetic field slowly changes towards the vertical direction.
Furthermore, the white arrows indicate that the magnetic field is weakly pinched towards the center.
On the other hand, the bottom left panel of figure \ref{beta_xz_small_MRN} show that the contours of
plasma $\beta$ in model\_MRN are dense around the midplane, and
the white arrows show that the magnetic field is strongly pinched towards the center.
This difference comes from the strength of ambipolar diffusion
and significant magnetic field drift in the pseudo-disk of model\_a03$\mum$.

Figure \ref{density_xy_small_a03mum} shows the density evolution on the $x$-$y$ plane.
In this model, the circumstellar disk also forms immediately after protostar formation
and survives for $\sim 10^4$ yr after protostar formation.
At its formation epoch, the disk size is similar to that of model\_MRN.
As it grows, the radius becomes larger than that of model\_MRN.
This is particularly clear in later epochs (bottom panels).
The spiral arms are also more prominent in this model.
We can clearly see that the disk monotonically grows in this model.

\subsection{Magnetic field drift induced by ambipolar diffusion}
\label{B_drift}
One of the most important phenomena caused by ambipolar diffusion
is magnetic field drift in the envelope.
Magnetic field drift determines the magnetic field strength
of a newly born circumstellar disk.
It has also been suggested as a mechanism for magnetic
flux redistribution in the envelope \citep[e.g.,][]{1998ApJ...497..850L}.
Furthermore, ion-neutral drift, which accompanies magnetic field
drift in the low-density region, may
provide possible direct observational evidence of a
relatively strong magnetic field in the
envelope \citep[see recent attempt of][]{2018A&A...615A..58Y}.

\begin{figure}
  \includegraphics[clip,trim=0mm 0mm 0mm 0mm,width=50mm,,angle=-90]{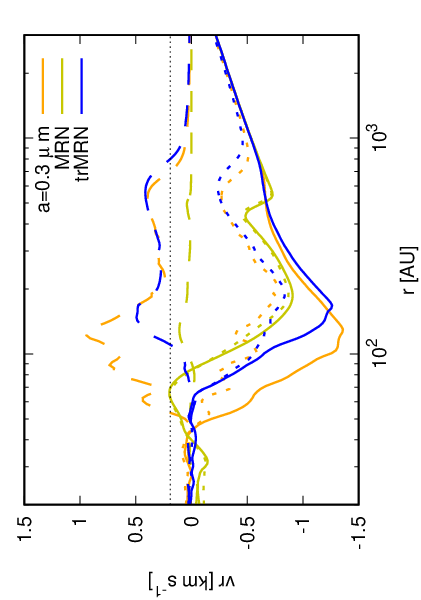}
\caption{
  The azimuthally averaged radial velocity on the $x$-$y$ plane as a function of radius
  at $t=5.3\times 10^4$ yr ($\sim 1.1\times 10^4$ yr after protostar formation).
  The orange, blue, and yellow lines show the results of model\_a03$\mum$, model\_trMRN,
  and model\_MRN, respectively.
  The solid, dashed, and dotted lines indicates the radial velocity of gas $v_{\rm r}$, drift velocity of magnetic field $v_{\rm drift, r}$,
  and $v_{\rm r}+v_{\rm drift, r}$, respectively.
  The thin dotted line shows $v=0.19 \kms$, which roughly corresponds to the sound speed at $T=10$ K.
}
\label{vdrift_comp}
\end{figure}

Figure \ref{vdrift_comp} shows the azimuthally averaged radial velocities
on the $x$-$y$ plane 
at $t=5.3\times 10^4$ yr ($\sim 1.1 \times 10^4$ yr after protostar formation).
The solid, dashed, and dotted lines show the gas radial velocity
$v_{\rm r}$, radial drift velocity of magnetic field $v_{\rm drfit, r}$, and
radial velocity of magnetic field $v_{\rm drfit, r }+v_{\rm r}$, respectively.
The drift velocity is calculated as
\begin{eqnarray}
\vel_{\rm drift} \equiv \eta_A \frac{(\nabla \times \magB) \times \magB}{|\magB|^2}.
\end{eqnarray}

  Among our simulations, model\_a03$\mum$  and model\_trMRN 
  show significant radial drift in relatively extended region with size of $r>100$ AU.
  In figure \ref{vdrift_comp}, we also plot $v_{\rm dirft, r}$ of model\_MRN 
  as an example of the model with small drift velocity.
  
  In model\_a03$\mum$, the drift velocity is positive
  and reaches $\sim 1 \kms$ at $r\sim 100$ AU (dashed line).
  As a result, total radial velocity of
  the magnetic field ($v_{\rm r}+v_{\rm drift, r}$)
  is $\sim -0.5 \kms$ (dotted line) and much
  slower than the gas infall velocity (solid line).
  We define $r_{\rm drift}$ as the maximum value of radius at which the radial magnetic
  field drift velocity is larger than $0.19 \kms$ on $x$-$y$ plane.
  Here we use velocity threshold of $0.19 \kms$ which corresponds to 
  the sound velocity of $T=10$ K and approximates the sound velocity in envelope.
  In model\_a03$\mum$,  $r_{\rm drift}$ is $\sim 700$ AU at $t=5.3\times 10^4$ yr.
  
  The radial drift velocity becomes small as the typical dust size decreases.
  In model\_trMRN, the drift velocity is typically $\sim 0.4 \kms$.
  On the other hand, $r_{\rm drift}$ also extends to $\sim 700$ AU in this model.
  In model\_MRN, notable outward radial drift can not be
  observed, and $v_{\rm r}+v_{\rm drift, r}$ (dotted line) is almost
  identical to the gas infall velocity ($v_{\rm r}$: solid line).
  This clearly shows that outward radial drift of the magnetic field occurs
  only with relatively large dust grains (or absence of small dust grains)
  in early evolution phase of young stellar objects.
  This is mainly because the difference of $\eta_A$
  in the low density region of $\rho \lesssim 10^{-14} \gcm$.

One of the most important aspects of magnetic field drift
by ambipolar diffusion is that magnetic field drift accompanies ion-neutral drift.
The ion-neutral drift is possibly observable as a velocity difference between
for example CO and HCO$^+$.
\citet{2002ApJ...573..199N} showed that  the ratio of drift
velocity of ion ($v_{\rm drift,ion}$) and
magnetic field ($v_{\rm drift}$) can be calculated as
\begin{eqnarray}
  \frac{v_{\rm drift,ion}}{v_{\rm drift}} = \frac{\beta_{\rm Hall}^2}{(1+\beta_{\rm Hall}^2)}(1+\xi \beta_{\rm Hall}^{-1}),
\end{eqnarray}
where $\xi$ is a constant of order unity, and  $\beta_{\rm Hall}$ is the Hall parameter,
\begin{eqnarray}
\beta_{\rm Hall} \equiv \frac{q_i B}{m_i c}\frac{m_i+m_{n}}{\rho \langle \sigma v \rangle_i},
\end{eqnarray}
where $q_i $ and $m_i$ are the charge and mass of the ion, respectively.
$m_{n}$ is the mean neutral mass.
$\langle \sigma v \rangle_i$ is the rate coefficient for collisional
momentum transfer between ion and neutral.
For the ion species with $\beta_{\rm Hall} \gg 1$,
the ratio of velocity becomes $v_{\rm drift,ion}/v_{\rm drift} \sim 1$ and
the ion moves with the magnetic field, giving ion-neutral relative velocities.

\begin{figure}
  \includegraphics[clip,trim=0mm 0mm 0mm 0mm,width=50mm,,angle=-90]{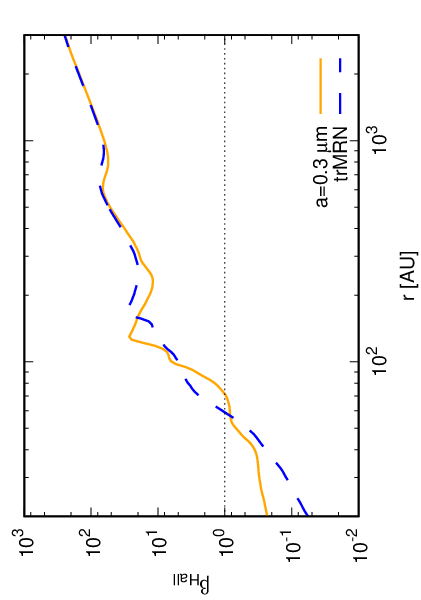}
\caption{
  The azimuthally averaged Hall parameter on the $x$-$y$ plane as a function of radius.
  The orange solid and blue dashed lines show the results of model\_a03$\mum$
  and model\_trMRN $t=5.3 \times 10^4$ yr.
}
\label{r_beta_hall}
\end{figure}

Figure \ref{r_beta_hall} shows the azimuthally
averaged Hall parameter on the $x$-$y$ plane of model\_a03$\mum$
and of model\_trMRN at $t=5.3\times 10^4$ yr.
Here, we assume that HCO$^+$ is a major ion species
in the envelope, and we use its mass and charge to calculate the Hall parameter.
Furthermore, in this figure, we assume $\langle \sigma v \rangle_i $ is constant and
$\langle \sigma v \rangle_i =1.8 \times10^{-9} {\rm cm^3 ~s^{-1}}$, which is enough for our purpose here.
Note that the Hall parameter of other ion species is not significantly different
because the mass and charge of the ion species in the envelope
do not differ significantly.
Figure \ref{r_beta_hall} shows that Hall parameter is $\beta_{\rm Hall} > 1$
in the region of $r\gtrsim100$ AU,
and the ions are expected to move with the magnetic field in the envelope.
Therefore, we conclude that the ion is well coupled with the magnetic field in the flattened envelope
and the magnetic-field drift velocity in figure \ref{vdrift_comp} can also be regarded as
the ion drift velocity in $r\gtrsim100$ AU.

\begin{figure}
  \includegraphics[clip,trim=0mm 0mm 0mm 0mm,width=50mm,,angle=-90]{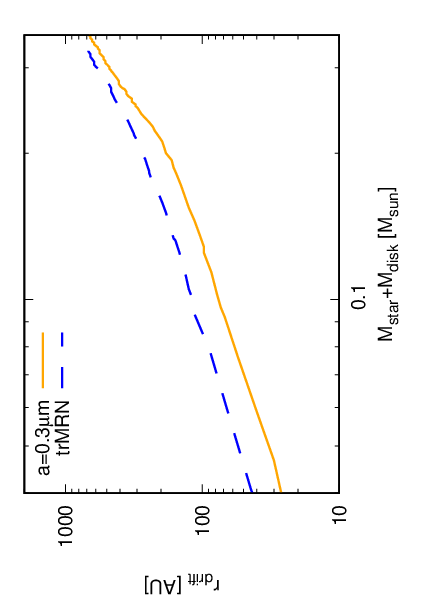}
\caption{
  Time evolution of $r_{\rm drift}$ (the radius of region with $v_{\rm drift, r}>0.19 \kms$)
  as a function of $M_{\rm star}+M_{\rm disk}$.
  The orange solid and blue dashed line show the result of 
  model\_a03$\mum$ and model\_trMRN, respectively.
}
\label{rdrift}
\end{figure}

  The size of the region in which ion-neutral drift occurs
  is important quantity to observe ion-neutral drift.
  Figure \ref{rdrift} shows the time evolution of $r_{\rm drift}$
  as a function of $M_{\rm star}+M_{\rm disk}$  for model\_a03$\mum$ and for model\_trMRN.
  Here $M_{\rm star}$ is the mass of the sink particle and $M_{\rm disk}$ is the disk mass
  which is defined as enclosed mass of the region
  with $\rho>10^{-14} \gcm$ (see \S \ref{disk_mass_eovolution}).
  The figure shows that $r_{\rm drift}$ monotonically
  increases as the mass (and hence magnetic flux) accumulates
  to the central region and 
  reaches $\sim 100$ AU at $M_{\rm star}+M_{\rm disk} \sim 0.1 \msun$.
  This indicates that $M_{\rm star}+M_{\rm disk} \gtrsim 0.1 \msun$
  is required for magnetic field drift to occur in $r>100$ AU and 
  indicates that a sufficient amount of mass (and hence magnetic flux)
  should have accreted to the central region for magnetic field drift
  to occur in a relatively extended region.
  We discuss the interpretation of our results
  and the relation to the observations in \S \ref{discussion_in_drift}.



\subsection{Warp of pseudo-disk}
\label{warp}
As shown in the figures \ref{density_xz_small_MRN} and \ref{beta_xz_small_MRN},
an interesting structure develops in the pseudo-disk
i.e., the warp of the pseudo-disk. Among our simulations, the warp appears
in model\_MRN,  model\_a01$\mum$
and model\_a0035$\mum$ at $\sim 10^4$ yr after protostar formation.

\begin{figure*}
\includegraphics[width=100mm,,angle=-90]{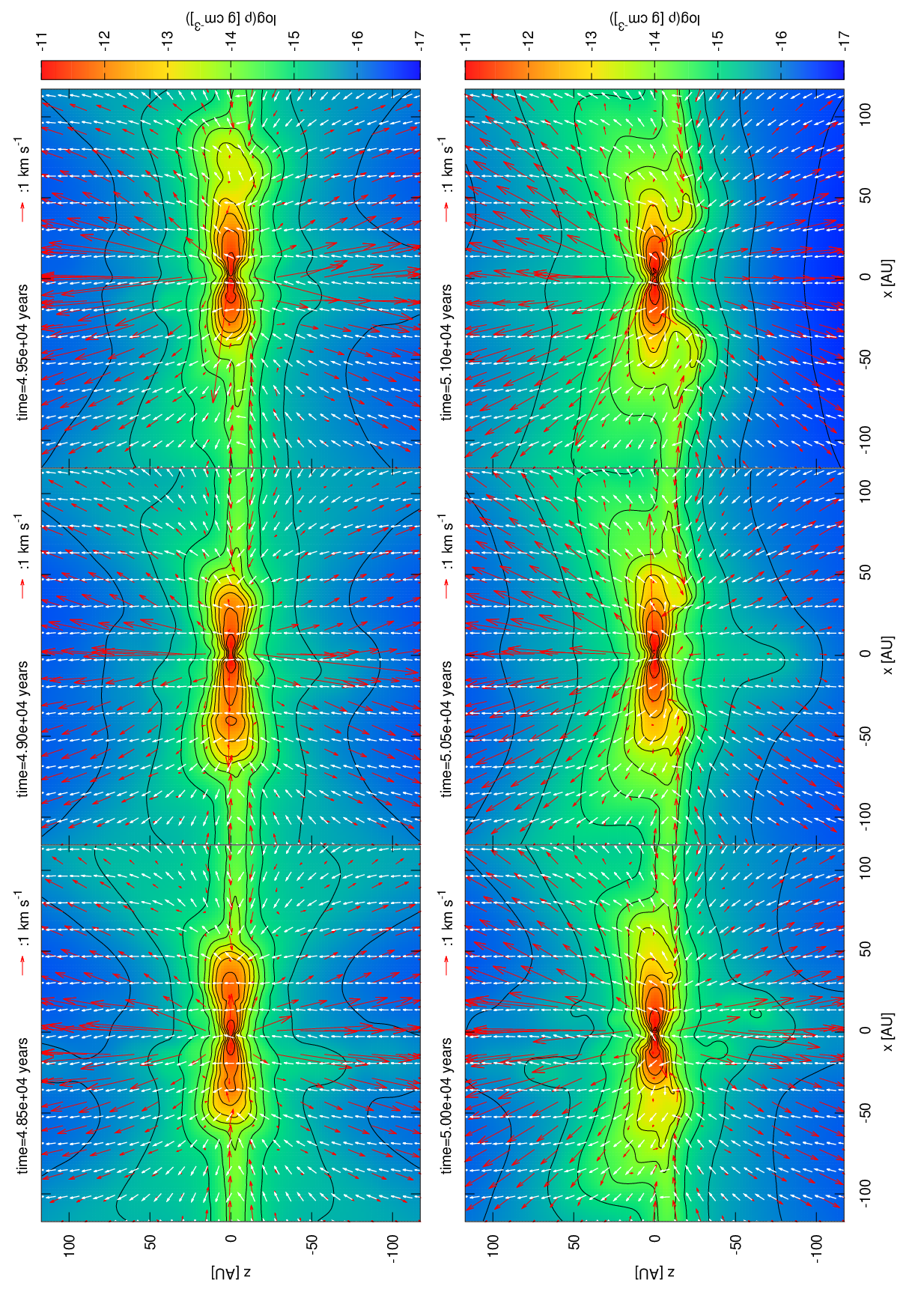}
\caption{
Density cross-sections on the $x$-$z$ plane for
central 500-AU square region during the development of the warp of the pseudo-disk
of model\_a01$\mum$.
The red arrows show the velocity field, and white arrows indicate the direction of magnetic field.
}
\label{warp_evolution}
\end{figure*}

Figure \ref{warp_evolution} shows 
time evolution of density for model\_a01$\mum$ from $t=4.85\times 10^4$ yr
(just before the warp develops) to $t=5.1 \times 10^4$ yr
and that how warp develops.
At $t \sim 4.85 \times 10^4$ yr (top left panel),
the density structure is approximately symmetric with respect to the $x$ axis.
Then, the magnetic field around the disk is perturbed by the spiral arm,
and the neck of the hour-glass magnetic field
is shifted to $z<0$ direction (top right panel). The gas accretion
flow is also shifted towards $z<0$
direction (bottom middle panel)
and the neck contracts and the warp develops.
The warp becomes prominent at $t \sim 5.1\times 10^4$ yr.
 Note that the outflow velocity becomes asymmetric and larger in $z>0$ region
  as the warp develops.

  Similar structures have been obtained in some previous studies.
  \citet{2013ApJ...763....6T} showed that the warp of pseudo-disk forms
  in their ideal MHD simulations (although their morphology is
  asymmetric with respect to $z$ axis).
  \citet{2016MNRAS.457.1037W} and \citet{2018MNRAS.473.4868Z} also reported the warps
  which are very similar to those obtained in our simulations.
  In turbulent cloud cores, the pseudo-disk tends to have more complicated
  warp structure as reported by \citet{2019MNRAS.489.5326L}.
  Note also that \citet{2003ApJ...591L.119L} analytically showed that
  disk-like structure with hour-glass magnetic field can be unstable.
  Although they investigate the instability of accretion disk
  and their results can not directly be applied to our configuration,
  the key mechanism is that the perturbation towards the vertical direction
  can grow by the Lorentz force.
  Thus, we speculate the warp structure may be related to this instability.
  In Appendix \ref{appendix_warp}, we show the results of the numerical tests 
  to reinforce the notion that the warp has physical origin.

\begin{figure}
  \includegraphics[width=50mm,,angle=-90]{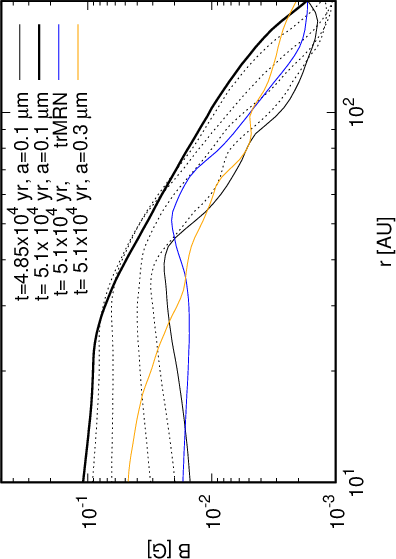}
  \caption{
    The azimuthally and vertically averaged magnetic field strength $|\magB|$
    of model\_a01$\mum$, model\_a03$\mum$ and model\_trMRN.
    as a function of radius during warp development.
    The vertical average is performed in $|z|<50$ AU.
    The two black solid lines show $|\magB|$
    at $t=4.85\times 10^4 $ yr (thin line) and  $5.1\times 10^4$ yr
    (thick line), and the black dashed
    line show $|\magB|$ at $t=4.9\times 10^4 $ yr,  $t=4.95\times 10^4 $  yr,
    $t=5.0\times 10^4 $  yr, $t=5.05\times 10^4 $  yr which correspond to the epochs
    shown in figure \ref{warp_evolution}.
    The blue and orange solid lines show $|\magB|$
    of model\_trMRN, and model\_a03$\mum$ at $t=5.1\times 10^4 $ yr, respectively.
}
\label{warp_B}
\end{figure}

The warp of the pseudo-disk has a negative impact on disk growth.
Due to the warp, the "neck" of the hour-glass magnetic field
contracts (see white arrows in figure \ref{warp_evolution}) and magnetic flux
in the disk increases.
As a result, the magnetic field is strengthened in the disk.
Figure \ref{warp_B} shows that magnetic field strength at $r\sim 10$ AU
increases from $10^{-2} $ G to $10^{-1}$ G  during warp development.
The magnetic field is vertically density weighted averaged in $|z|<50$ AU and
is also azimuthally averaged.
The magnetic field with warp 
becomes stronger than that without warp at the corresponding epoch.
The increase of the magnetic field strength by the warp enhances the magnetic
braking in the disk, and the disk angular momentum
and disk size begin to decrease (see figures \ref{disk_J} and
\ref{density_xy_small_MRN}) after warp formation.
Our results show that there is an evolution path in
which the disk begins to shrink $\sim 10^4$ yr after its birth.

In our simulations, the warp is formed
in model\_MRN, model\_01$\mum$, and model\_a0035$\mum$, 
and the pseudo-disks in other simulations
are approximately symmetric with respect to the $x$-$y$ plane
at least within $ \lesssim 10^4$ yr after protostar formation.
Whether the warp of the pseudo-disk develops in other models in subsequent
(long-term) evolution and the detailed physics triggering its formation
are unclear and further study is required.

\subsection{Time evolution of angular momentum and disk radius}
\label{disk_evolution}
In this subsection, we investigate the time evolution of angular momentum
and centrifugal radius of the central region to quantitatively
discuss disk size evolution after protostar formation.

In this paper, we do not use the disk criteria such as 
those considering rotation velocity and infall
velocity \citep{2011PASJ...63..555M,2016A&A...587A..32M}
or those considering gravitational force and centrifugal force
\citep{2015ApJ...810L..26T} because 
we found that they overestimate disk size due to contamination
from marginally outflowing region
when the structure of the inner envelope is elongated due to pseudo-disk warp.

The bottom right panels of figure \ref{density_xz_small_MRN},
\ref{beta_xz_small_MRN}, and \ref{density_xy_small_MRN} highlight this concern.
The bottom right panel of figure \ref{density_xy_small_MRN} shows that
there is a rapidly rotating low-density region of $r\sim 100$ AU on the $x$-$y$ plane
around the central disk. However, from the $x$-$z$ map of density and plasma
$\beta$ (bottom right panels of figure \ref{density_xz_small_MRN}
and figure \ref{beta_xz_small_MRN}),
such low-density and low $\beta$
regions at $r \sim 100$ AU on the $x$ axis do not match with our
intuition for the circumstellar disk.
Furthermore, as we will show later,
the angular momentum of the region decreases
after the formation of the warped pseudo-disk,
although the disk size estimated by the above-mentioned criteria increases.
This strongly suggests that the criteria for disk are
inappropriate when the system loses its symmetric structure.

Instead, we use total angular momentum and centrifugal
radius of the central region to estimate disk size,
which is more physically rigorous and free from
disk size overestimation.
The angular momentum of disk $J(\rho_{\rm disk})$ is calculated as
\begin{eqnarray}
J(\rho_{\rm disk})\equiv \left| \int_{\rho>\rho_{\rm disk}} \rho (\rad \times \vel) d V \right|.
\end{eqnarray}
For the density threshold of the disk, we choose $\rho_{\rm disk}=10^{-14} \gcm$.
We confirmed that our results below do not strongly depend on the choice of $\rho_{\rm disk}$
(see also, contours of figure \ref{density_xy_small_MRN} and \ref{density_xy_small_a03mum}).
The centrifugal radius is then calculated as
\begin{eqnarray}
r_{\rm disk} \equiv  r_{\rm cent}=  \frac{\bar{j}(\rho_{\rm disk})^2}{G M_{\rm star}}.
\end{eqnarray}
Here $\bar{j}(\rho_{\rm disk})=J(\rho_{\rm disk})/M(\rho_{\rm disk})$
where $M(\rho_{\rm disk})$ is the enclosed mass within the region $\rho>\rho_{\rm disk}$.
We regard this centrifugal radius as a disk radius.

The solid lines of figure \ref{disk_J} show time evolution of the disk radius.
The disk radius at the protostar formation epoch is $r \sim 10$ AU and 
increases in all simulations up to $t \lesssim 5\times 10^4$ yr.
However, it begins to decreases in 
model\_a0035$\mum$ (red), model\_a01$\mum$ (black) and model\_MRN (yellow).
The epochs of disk size decreases corresponds to the epochs of the warp formation.
As shown in figure \ref{warp_B}, the warp of the pseudo-disk strengthens
the magnetic field (and magnetic torque) in the disk
and has a negative impact on disk growth.
In model\_a03$\mum$ and model\_trMRN, the disk radius as well as $J(\rho_{\rm disk})$
continues to increase until the end of the simulation.
Among the models, the maximum disk radius of $r\sim 60$ AU is
realized in model\_a03$\mum$.
The disk size (and angular momentum) evolution of model\_trMRN are almost
identical to those of model\_a03$\mum$.
The evolution of disk radius and angular momentum shows
that whether the warp occurs or not affects disk size evolution.


%


\subsection{Time evolution of mass of protostar and disk}
\label{disk_mass_eovolution}
Figure \ref{disk_mass} shows the time evolution of
disk mass $M_{\rm disk}$ (solid) and protostellar mass $M_{\rm star}$ (dashed).
For disk mass, we used enclosed mass within $\rho>\rho_{\rm disk}=10^{-14} \gcm$.
Again, we confirmed that the disk mass does not strongly depends on $\rho_{\rm disk}$, and
$\rho_{\rm disk}=10^{-13}\gcm$ and $\rho_{\rm disk}=10^{-15}\gcm$  give almost identical results.
For the stellar mass, we plot the mass of the sink particle.

At the protostar formation epoch, the disk mass
is $M_{\rm disk}\sim 4 \times 10^{-2} \msun$, which
corresponds to the mass of the pressure-supported
first core \citep{1969MNRAS.145..271L,1998ApJ...495..346M}, and
$M_{\rm disk}$  does not significantly decrease around $t=4.2\times 10^4$ yr.
Thus, the most part of the first core does not accrete onto the central protostar but stays around the
protostar. This means that most of the gas in the first core is directly transformed into the disk.
This formation picture of the circumstellar disk from the first core was
suggested by \citet{2011MNRAS.413.2767M} and \citet{2012PTEP.2012aA307I},
and our results are consistent with theirs.

After protostar formation, disk and protostellar mass increase almost monotonically and 
reach $\sim 0.1\msun$ within $10^4$ yr after protostar formation.
The slight decrease of disk mass in model\_a01$\mum$
and model\_MRN is due to enhancement of magnetic braking by the warp of the pseudo-disk.
The mass of the disk is comparable to or larger than the protostellar mass
within $10^4$ yr after protostar formation (or until the central star mass becomes $\sim 0.1 \msun$). 
We discuss that a massive disk is a natural
consequence of large mass accretion from the envelope in \S \ref{discussion_disk_formation}.

The total mass in the central region is slightly different 
among the models ($\sim 5 \times 10^{-2} \msun$ at $t\sim 5.3 \times 10^4$ yr, for example).
This difference in total mass is consistent with the
difference in outflow mass.  Mass removal by outflow causes the
difference of the total mass in the center.

\subsection{Time evolution of mass accretion rate}
In this subsection, we investigate mass accretion rate, which is
a fundamental parameter for the evolution of YSOs.
The envelope-to-disk mass accretion rate is calculated as
$\dot{M}_{\rm env, disk}=\Delta M_{\rm enclose}/\Delta t_{\rm interval}$,
where $\Delta M_{\rm enclose}$ is the difference 
of the mass within the region of $\rho>10^{-14} \gcm$
(including the mass of the central protostar) 
during the interval of $[t,t+\Delta  t_{\rm interval}]$.
$\Delta t_{\rm interval}$ is chosen to be $3 \times 10^{2}$ yr.
the results were found to be almost unchanged
with the other value of $\Delta t_{\rm interval}$ and
our results discussed below barely depends on the choice of $\Delta t_{\rm interval}$.
The disk-to-star mass accretion rate is calculated as
$\dot{M}_{\rm disk, star}=\Delta M_{\rm star}/\Delta  t_{\rm interval}$,
where $\Delta M_{\rm star}$ is the difference in the mass of the protostar
during the interval $[t,t+\Delta  t_{\rm interval}]$.

Figure \ref{disk_mdot} shows the mass accretion rates.
The solid lines show the mass accretion rate from disk to protostar $\dot{M}_{\rm disk, star}$.
The mean value during the evolution is $\dot{M}_{\rm disk, star} \sim 10^{-5} \msun {\rm year}^{-1}$.
This mass accretion rate corresponds to almost the upper limit of the
observed mass accretion rate of Class 0 YSOs \citep{2017ApJ...834..178Y}.
The temporal oscillation of  $\dot{M}_{\rm disk, star}$ is due to mass accretion by the spiral arms.
The amplitude of the oscillation is a factor of two to three.
The dashed lines show envelope-to-disk mass accretion rate and show
that  $\dot{M}_{\rm env, disk} \sim 4 \times 10^{-5} \msun {\rm year}^{-1} $
at the protostellar formation epoch and decreases with time.
At the end of the simulation,
it decreases to $\dot{M}_{\rm env, disk} \sim 10^{-5} \msun {\rm year}^{-1} $.
The oscillation of $\dot{M}_{\rm env, disk}$ in the latter phase is due to density
fluctuation of the outer disk by the spiral arms or warp and
whether the oscillation appears or not  depends on the choice of the density threshold.
Thus, we think the oscillation does not reflect the real change of accretion rate
towards the central region.
The mass accretion rate of $\dot{M}_{\rm env, disk} \sim  4 \times 10^{-5} \msun {\rm year}^{-1}$
at the protostar formation epoch and its decrease in the latter accretion phase
are quantitatively consistent with previous analytic and simulation studies
of dynamical collapse of cloud cores
\citep{1985MNRAS.214....1W,1998ApJ...493..342S, 1996PASJ...48L..97T,
  2005MNRAS.360..675V}.
Note that the accretion rate of a dynamical collapse becomes
much larger than that of a collapse of a singular isothermal sphere 
$\dot{M}_{\rm env, disk} \sim  2 \times 10^{-6} \msun {\rm year}^{-1}$ \citep{1977ApJ...214..488S}
because of larger density and larger infall velocity of the envelope.



\begin{figure}
   \includegraphics[width=50mm,,angle=-90]{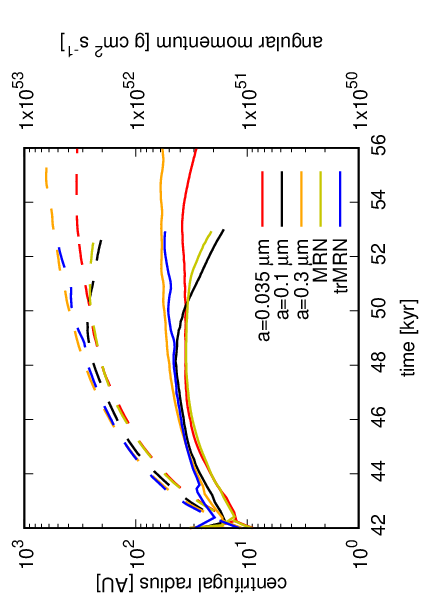}
\caption{
  Time evolution of  centrifugal radius (solid lines; left axis) and
  the angular momentum (dashed lines; right axis).
  The red, black, orange, blue, and yellow lines corresponds to
  model\_a0035$\mum$, model\_a01$\mum$, model\_a03$\mum$,
  model\_MRN, and model\_trMRN, respectively.
}
\label{disk_J}
\end{figure}

\begin{figure}
   \includegraphics[width=50mm,,angle=-90]{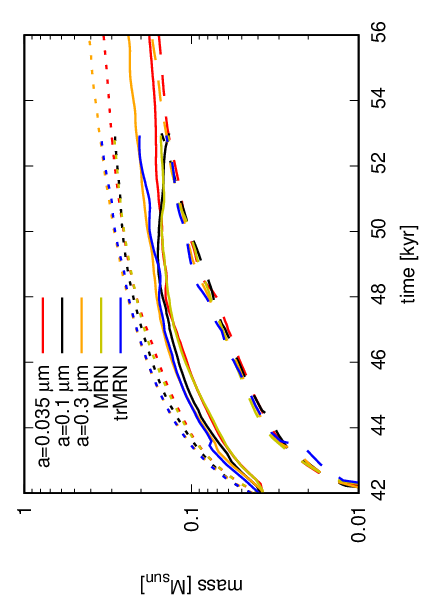}
\caption{
  Time evolution of mass of disk $M_{\rm disk}$ (solid lines), protostar $M_{\rm star}$ (dashed), and $M_{\rm star}+M_{\rm disk}$ (dotted).
  The red, black, orange, blue, and yellow lines show the results of model\_a0035$\mum$, model\_a01$\mum$, model\_a03$\mum$,
  model\_MRN, and model\_trMRN, respectively.
}
\label{disk_mass}
\end{figure}

\begin{figure}
   \includegraphics[width=50mm,,angle=-90]{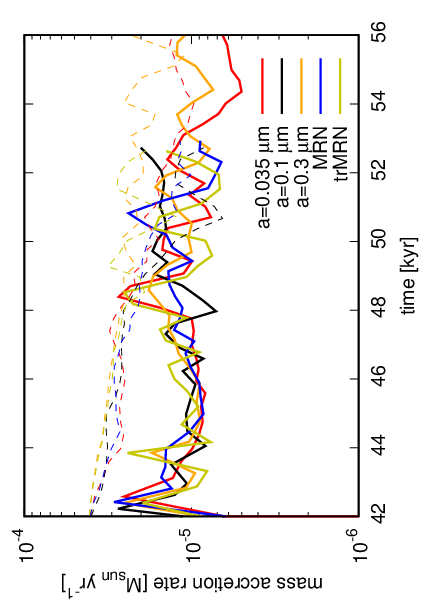}
\caption{
  Time evolution of mass accretion rate. The solid lines show the disk-to-star mass accretion rate $\dot{M}_{\rm disk, star}$ and
  dashed lines show the envelope-to-disk mass accretion rate $\dot{M}_{\rm env, disk}$.
  The red, black, orange, blue, and yellow lines show the results of model\_a0035$\mum$, model\_a01$\mum$, model\_a03$\mum$,
  model\_MRN, and model\_trMRN, respectively.
}
\label{disk_mdot}
\end{figure}

\subsection{Time evolution of outflows}
\label{outflow_evolution}
In this subsection, we investigate the properties of outflows.
We define an outflow as a region
that satisfies $v_{\rm r} = (\vel \cdot \rad)/|\rad|>2 c_{\rm s,iso}$ and $\rho<10^{-14} \gcm$
where $c_{\rm s,iso}=0.19\kms$ is sound velocity at $T= 10$ K.
Outflows are formed in all models.

\begin{figure*}
  \includegraphics[width=50mm,,angle=-90]{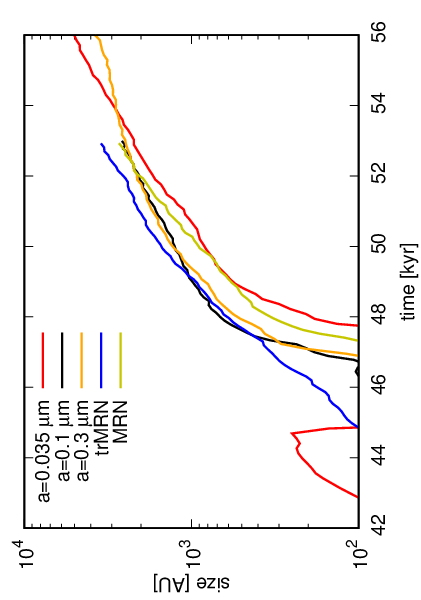}
  \includegraphics[width=50mm,,angle=-90]{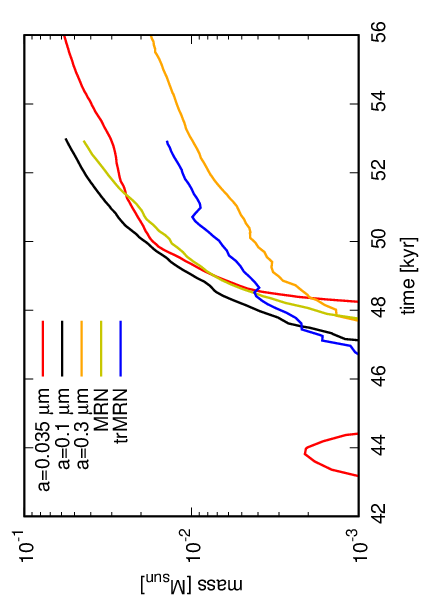}
  \includegraphics[width=50mm,,angle=-90]{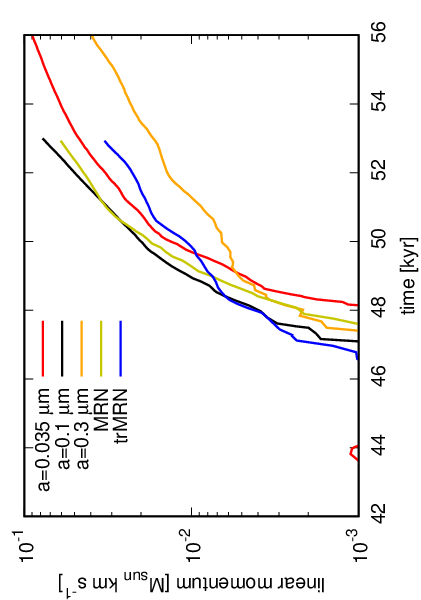}
  \includegraphics[width=50mm,,angle=-90]{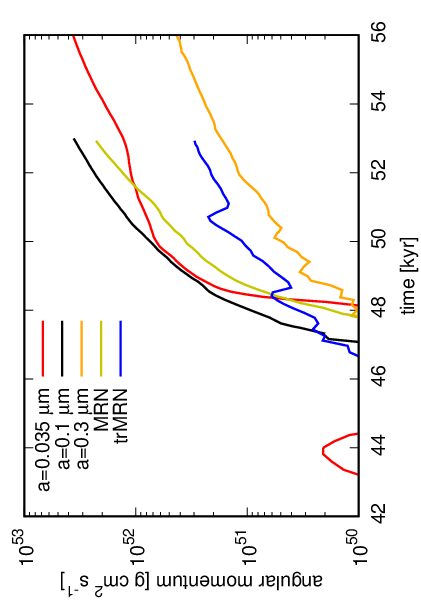}
\caption{
  Time evolution of size (top left), mass (top right), linear momentum (bottom left), and angular momentum (bottom right) of outflow.
  The red, black, orange, blue, and yellow lines show the results of model\_a0035$\mum$, model\_a01$\mum$, model\_a03$\mum$,
  model\_MRN, and model\_trMRN, respectively.
}
\label{outflow}
\end{figure*}

The top left panel of figure \ref{outflow} shows time evolution of outflow size.
The outflow size is defined as the distance of the particle in the outflow
that is farthest from the central protostar.
In our simulations, strong outflows with velocity of $v>1 \kms$ form 
several thousand years after protostar formation
(note that the protostars form at $t=4.15\times 10^4$ yr).
Outflow size monotonically increases and reaches $\sim 3000$ AU
in $\sim 5\times 10^3$ yr after outflow formation.
This suggests that the mean velocity of the outflow head is $\sim 3 \kms$.

The top right panel of figure \ref{outflow} shows the time evolution of the outflow mass $M_{\rm out}$.
$M_{\rm out}$ monotonically increases in all models.
The outflow mass is anti-correlated with the disk size,
suggesting that the outflow activity is related to disk growth 
The outflow masses and dynamical timescales obtained in our simulations
are consistent with observed outflows with dynamical timescale
$t_{\rm dyn}<10^4$ yr, whose mass
ranges from $10^{-2} \msun$ to $10^{-1} \msun$ \citep{2004A&A...426..503W}.


The bottom left panel shows the linear momentum of outflow $P_{\rm out}$.
The difference in linear momentum mainly comes from the difference of mass of outflow, and
the mean velocity of outflow $\equiv P_{\rm out}/M_{\rm out}$ is $\sim 2 \kms$ in all simulations.
This estimate is consistent with the outflow velocity estimated from the size and the age.

The bottom right panel shows the outflow angular momentum.
The outflow angular momentum also shows anti-correlation to the disk size
(compare this panel with figure \ref{disk_J})
which is consistent with \citet{2016MNRAS.457.1037W}.
The difference in outflow angular momentum among the models
is an order of magnitude (e.g., at $t\sim 5.3 \times 10^4$ yr)
and does not merely come from the difference in mass.
The outflow angular momentum is comparable to or even larger than
disk angular momentum in $t\gtrsim 5\times 10^4$ yr for model\_a0035$\mum$,
model\_a01$\mum$, and model\_MRN.
Coincidently, the disk size begins to decrease at this epoch.
Thus, the angular momentum removal due to outflow may play
a role in disk size evolution for these models.
Note also that pseudo-disk warping occurs in these models.
On the other hand, in model\_a03$\mum$ and  model\_trMRN, the outflow angular momentum
remains smaller than the disk angular momentum.
This indicates that angular momentum removal by outflow plays a
minor role for disk evolution in these models.


\section{Discussion}
\label{discussion}
\subsection{Formation and early evolution of circumstellar disk}
\label{discussion_disk_formation}
\subsubsection{Magnetic braking}
In all models considered in this paper, circumstellar disks are formed immediately
after protostar formation.
The mass and size of the circumstellar disk are $\sim 10$ AU
and $\sim 4\times 10^{-2} \msun$ at its formation epoch, respectively.
The initial size and mass are largely consistent with those
of the first core \citep{1969MNRAS.145..271L,1998ApJ...495..346M},
indicating  that the first core directly transforms into
the circumstellar disk \citep{2011MNRAS.413.2767M,2012PTEP.2012aA307I}.
The disk grows to several tens of AU at $10^4$ yr after
protostar formation (protostellar mass reaches $M\sim 0.1 \msun$ at this epoch).

The magnetic braking catastrophe \citep{2008ApJ...681.1356M}, which claims
that disk formation is completely suppressed by magnetic braking in
an early phase of protostar evolution,
has been a long-standing issue in theoretical
studies on protostar formation \citep[e.g.,][]{
  2003ApJ...599..363A,2007MNRAS.377...77P,
  2008ApJ...681.1356M,2008A&A...477....9H,
  2009A&A...506L..29H,2012A&A...543A.128J,
  2012ApJ...747...21S,2013MNRAS.432.3320S,
  2013A&A...554A..17J,2013ApJ...774...82L,
  2011MNRAS.413.2767M,
  2011ApJ...733...54K,
  2011ApJ...738..180L,2013ApJ...763....6T,
  2015ApJ...801..117T,
  2015MNRAS.452..278T,2015ApJ...810L..26T,2016A&A...587A..32M,
  2016MNRAS.457.1037W,2018MNRAS.473.4868Z,2019MNRAS.489.5326L}.
The key physical mechanisms of magnetic braking catastrophe are 
magnetic flux freezing and a rapid increase in the magnetic
field towards the central star.
Magnetic diffusion is hence the most promising mechanism to overcome catastrophic magnetic braking because 
it relaxes the heart of the magnetic braking catastrophe,
i.e., the flux freezing between the magnetic field and gas.
Therefore, in principle, the magnetic braking catastrophe
can be solved by magnetic diffusion.
Our previous studies with three dimensional non-ideal MHD simulations have shown that
disk formation  (with size of $\sim 1$ to $10$ AU)
actually becomes possible immediately after
protostar formation by ohmic and ambipolar diffusion \citep[see the
comparison between ideal and resistive MHD simulation in][]{2015MNRAS.452..278T}.
The angular momentum shown in figure 6 of \citet{2015MNRAS.452..278T}
  is almost the same as that at the protostar formation epoch shown in figure \ref{disk_J}.
The simulations of our previous studies were, however,
halted just after protostar formation.
Therefore, whether the disk can survive and grow
after protostar formation was unclear.

The current study shows that the disk grows to several $10$ AU scale and 
is long-lived, at least for $\sim 10^{4}$ yr after protostar formation.
Recent theoretical studies considering magnetic diffusion have also confirmed 
that the disk is formed in a very early phase of protostar formation
\citep{2015ApJ...801..117T,2016MNRAS.457.1037W,2016A&A...587A..32M}.
Furthermore, many observational studies have shown that circumstellar disks
form in a very early phase of protostar formation
\citep[e.g.,][]{2013A&A...560A.103M,2014ApJ...796..131O,2017ApJ...834..178Y}. 
Thus, the statement that magnetic braking is catastrophic, in the sense that 
magnetic braking completely suppresses early disk formation, 
is not supported either theoretically or observationally.

However, note that the strength of magnetic braking depends on many factors,
such as initial conditions,
included physics, and microscopic chemistry.
Furthermore, the treatment of the central protostar
(or inner boundary condition) differs among theoretical studies.
This may cause quantitative differences in disk size evolution.
Thus, in some cases, a disk did not form even with the non-ideal MHD effect,
as indicated in \citet{2011ApJ...738..180L} and \citet{2018MNRAS.473.4868Z}
\citep[see also][for the impact of the inner boundary condition or sink]{2014MNRAS.438.2278M, 2020A&A...635A..67H}.
This is not surprising because of the differences in the other factors.
Note also that magnetic braking certainly has a negative impact on disk growth.
Compared to our previous studies on disk formation
of unmagnetized cloud core
\citep[e.g.,][]{2011MNRAS.416..591T,2013MNRAS.428.1321T,2013MNRAS.436.1667T,2015MNRAS.446.1175T},
the disk size is small and disk fragmentation is suppressed by the magnetic field.
For example, as shown in figure 1 of \citet{2011MNRAS.416..591T}, 
disk fragmentation is expected with our cloud
core ($\alpha_{\rm therm}=0.4$ and $\beta_{\rm rot}=0.03$) if the magnetic field is ignored.

\subsubsection{Impact of dust size on disk formation and evolution}
Recently, it has been suggested that the removal 
of small dust grains (or dust growth) enhances disk formation
\citep{2016MNRAS.460.2050Z,2018MNRAS.473.4868Z}. We have confirmed 
this conclusion.
As shown in figure \ref{disk_J}, 
an increase in dust size certainly enhances disk growth.
This is due to the enhancement of ambipolar
diffusion at $\rho\sim 10^{-14}\gcm$ (see figure \ref{eta_fig}).
However, we also found that the difference in
dust size does not qualitatively change the disk formation, i.e.,
it does not determine whether the disk forms or not.
Our numerical simulations showed that even with small 
dust grains, such as in model\_MRN or
model\_a0035$\mum$, the disk does form 
immediately after protostar formation and survives.

The occurrence of magnetic field drift in the envelope, on the other hand,
is different among the simulations with large and small dust grains.
This will change the magnetic field strength of the circumstellar disk by changing the
amount of brought-in magnetic flux to the disk.
The difference in magnetic field strength in the disk may affect the subsequent long-term
evolution of the disk (and possibly MRI activity in the disk).
Thus, longer-term simulations ($\gtrsim 10^5$ yr after protostar formation)
with various dust models would be an important subject for future study.

\subsubsection{Disk size evolution and comparison with the observations}
Recent observations have revealed that the circumstellar disk is formed 
in the early evolution phase of YSOs \citep[e.g.,][]{
  2013A&A...560A.103M,2014ApJ...796..131O,
  2015ApJ...812...27A,2017ApJ...849...56A,2017ApJ...834..178Y}.
Our results are qualitatively consistent with these results.
Then, are the simulation results quantitatively consistent with observations?

\begin{figure}
  \includegraphics[width=50mm,,angle=-90]{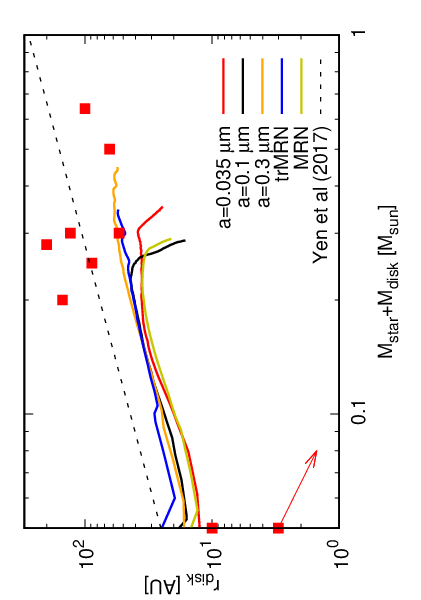}
\caption{
  Time evolution of disk size as a function of $M_{\rm star} + M_{\rm disk}$ and comparison with observations.
  The black, red, orange, blue, and yellow lines show the results of model\_a0035$\mum$, model\_a01$\mum$,model\_a03$\mum$,
  model\_MRN, and model\_trMRN, respectively.
  The red rectangles show the mass and disk size 
  of Class 0/I YSOs from  table 5 of \citet{2017ApJ...834..178Y} and \citet{2017ApJ...843...27L,2018ApJ...863...94L}.
  The names of objects are B335 ($M=0.05\msun$), HH211-mms ($M=0.05 \msun$), VLA1623 ($M=0.2\msun$),
  HH212 ($M=0.25\msun$), L1455 IRS1 ($M=0.28\msun$), L1527 IRS($M=0.3\msun$), Lupus 3 MMS ($M=0.3\msun$),
  L1551 IRS5 ($M=0.5\msun$), TMC-1A ($M=0.64\msun$) from left to
  right (for the objects with the same mass, bottom to top).
  The symbol with arrow indicates that the protostellar
  mass and disk size of the symbol are the lower and upper limit, respectively.
  The black dashed line shows the fitting formula of disk
  size of Class 0 YSOs from \citet{2017ApJ...834..178Y}.
}
\label{m_J}
\end{figure}

To answer this question, we plot the disk 
size of Class 0/I YSOs from \citet{2017ApJ...834..178Y}
and disk size evolution obtained in this study in figure \ref{m_J}.
The horizontal axis shows the sum of the disk and central star mass because
the mass of the protostar of the observations is estimated from the Keplerian rotation velocity at the
disk edge or infall velocity, and the contribution of the mass in the disk should also be included.
Figure \ref{m_J} shows that, in the late phase ($M>0.1 \msun$), our results with large dust grain size
are approximately consistent with the observations.
For example,  model\_a03$\mum$ (orange) 
and model\_trMRN (blue) have the almost same disk size of L1527 IRS.
On the other hand, the disk size in the simulations
with pseudo disk warp is generally 
smaller than the observational results.
However, note that the rotationally supported
(marginally outflowing) region extends to 100 AU in
these simulations (figure \ref{density_xy_small_MRN}) and this region
can be observationally regarded as rotationally supported disk,
and this possibly explains the discrepancy.
In earlier phase ($M<0.1 \msun$), disk size of the 
simulation tends to be larger than that of B335. 
This may be due to the difference of initial angular
momentum profile between our initial conditions and the conditions of 
the real cloud cores.
The early evolution phase of the disk is more sensitive 
to the initial angular momentum profile, and
a more realistic velocity field such as 
turbulence would be suitable to investigate early
phase disk evolution
\citep{2012ApJ...747...21S, 2013A&A...554A..17J,2017ApJ...839...69M,2018MNRAS.477.4241L,2019MNRAS.489.5326L,2020MNRAS.492.5641T}.


\subsubsection{Massive disk formation as a consequence of large mass accretion rate}
 Our results show that the disks formed in our simulations tends to be 
 massive enough to develop gravitational instability (or $Q \sim 1$)
where $Q$ is Toomre's $Q$ parameter.
Here after "massive disk" is used to 
mean the marginally gravitationally
unstable disk (or disk with $Q \sim 1$).
It is well known that a massive disk is  formed
in unmagnetized cloud cores \citep{1994ApJ...421..640N,
2003ApJ...595..913M,2006ApJ...650..956V,2009ApJ...704..715V,
2010ApJ...719.1896V,2010ApJ...724.1006M,
2011MNRAS.416..591T,2012MNRAS.427.1182S,
2013MNRAS.428.1321T,2013MNRAS.436.1667T,
2013ApJ...770...71T,2014MNRAS.439.3039L,2015MNRAS.446.1175T}.
Our results as well as recent theoretical studies have shown
that even with a magnetic field, disk becomes
massive and gravitationally unstable once it grows to several 10 AU
\citep{2011PASJ...63..555M,2015ApJ...810L..26T,2015MNRAS.452..278T,
  2016A&A...587A..32M,2018MNRAS.473.4868Z}.

The large mass accretion rate of $\gtrsim 10^{-6} \msunyear$ causes the
formation of the massive disk. 
This can be understood using the steady viscous accretion disk model as follows.
In the viscous accretion disk,
the mass accretion rate in the disk, temperature, and surface density,
and viscous parameter $\alpha$ are related to
\begin{eqnarray}
  \alpha \Sigma_{\rm gas} \frac{c_{\rm s}^2}{\Omega} = \frac{\dot{M}_{\rm disk}}{3 \pi},
\end{eqnarray}
where $\Sigma_{\rm gas}$,  $c_{\rm s}$, $\Omega$, and $\dot{M}_{\rm disk}$
are the gas surface density, sound velocity, orbital period,
and mass accretion rate of the circumstellar disk, respectively.
This can be rewritten as
\begin{eqnarray}
  \label{alpha_estimate}
\alpha &=& \frac{\dot{M}_{\rm disk} }{3  c_{\rm s}^3}\frac{c_{\rm s} \Omega}{\pi \Sigma_{\rm disk}} =\frac{1}{3}\frac{\dot{M}_{\rm disk}}{c_{\rm s}^3/G}Q   \nonumber \\
&=& 0.24 \left( \frac{\dot{M}_{\rm disk}}{3 \times 10^{-6} \msunyear} \right) 
 \left( \frac{Q}{2} \right) 
\left( \frac{T}{30 K} \right)^{-3/2},
\end{eqnarray}
where $Q$ is Toomre's Q parameter and we approximate
epicycle frequency $\kappa$ is $\kappa=\Omega$.
$\alpha$ is determined by disk temperature and the $Q$ value for 
a given mass accretion rate of disk.
This indicates that even with a massive disk
with $Q\sim 1$, $\alpha\sim 0.1 $ is required to
realize $\dot{M}_{\rm disk} \sim 3\times 10^{-6} \msunyear$,
which is expected value of mass accretion
rate in the early evolutionally phase of YSOs \citep[see, ][]{2017ApJ...834..178Y}.
If the disk is less massive (has large $Q$ value),
a large value of $\alpha$ is required to realize a mass accretion
rate of $\dot{M}_{\rm disk}\sim 3 \times 10^{-6} \msun {\rm yr^{-1}}$. 
For example, if $Q\sim 10$ in the early disk evolution stage,
$\alpha$ as large as $1.2$ is required.
However, $\alpha \gtrsim 1$ (meaning that trans- or super-sonic accretion)
as well as $T \gg 30$ K may be unrealistic 
for a disk with a size of several 10 AU to 100 AU around a low mass protostar.
One may think that the mass accretion rate in 
the disk is not necessarily radially constant and
high mass accretion rate only at the inner 
hot region of disk explains the mass accretion rate of the protostar.
However, in Class 0/I YSOs, the gas is continuously 
supplied from the envelope to the disk with 
a mass accretion rate of $\dot{M}_{\rm env, disk} \sim 10^{-6} \msun {\rm yr^{-1}}$.
Thus, if $\dot{M}_{\rm disk}$ varies in the disk
and $\dot{M}_{\rm disk}$ in the outer region of disk is
small, the gas stagnates in the outer region and disk mass increases 
due to envelope accretion.
This simple estimate suggests that a massive disk forms
when mass accretion rate of $10^{-6}$ to $10^{-5} \msunyear$
and disk size of several 10 AU to 100 AU are simultaneously realized.

The disk mass estimated from the observations of Class 0 YSOs
does not strongly contradict the mass of a marginally gravitationally unstable disk.
The disk surface density and disk mass with $Q\sim 2$ are estimated as
\begin{eqnarray}
  \Sigma_{\rm GI}(r) = 4.1  \left(\frac{M_{\rm star}}{0.1 \msun}\right)^{1/2} \nonumber \\ 
\left(\frac{Q}{2}\right)^{-1} \left(\frac{r}{100 {\rm AU}} \right)^{-12/7} {\rm g~ cm^{-2}}, \\
M_{\rm GI} = \int 2 \pi r \Sigma_{\rm GI} (r) dr 
\sim 1.0 \times 10^{-1} \left(\frac{M_{\rm star}}{0.1 \msun}\right)^{1/2}
\nonumber \\ 
\left(\frac{Q}{2}\right)^{-1} \left(\frac{r_{\rm out}}{100 {\rm AU}} \right)^{2/7} \left(1-\left(\frac{r_{\rm in}}{r_{\rm out}}\right)^{2/7}\right), \msun \nonumber \\
\end{eqnarray}
where we assume $T=150 ~(r/ 1 {\rm AU})^{-3/7} [{\rm K}]$ 
\citep{1970PThPh..44.1580K,1997ApJ...490..368C}, Keplerian rotation,
and $Q={\rm const}$ in $r_{\rm in}<r<r_{\rm out}$. $r_{\rm in}$ and $r_{\rm out}$ are 
the inner and outer radii of the gravitationally unstable 
region, respectively.
We set the $Q$ value for the marginally unstable disk to 
$Q_{\rm crit}=2$ because the spiral arms develop at $Q\sim 1.4$
\citep{1994ApJ...436..335L} and
the marginally unstable disk may
have a slightly larger value than $1.4$.
Thus, a gravitationally unstable disk 
with radius of $\sim 100$ AU has mass of $\sim 0.1 \msun$.

On the other hand, \citet{2009A&A...507..861J} estimated 
that the disk mass of Class 0 YSOs has a mean value of $\sim 0.05 \msun$ ranging from $0.01-0.46 \msun$.
\citet{2011ApJS..195...21E} estimated that the disk mass has a mean value of $\sim  0.2 \msun$ ranging from
$0.04-0.28 \msun$, except for one significantly massive disk. These two studies did not resolve the disks.
More recently, \citet{2018ApJ...866..161S} reported the disk mass of Class 0 YSOs ranging from $0.03 - 0.46 \msun$
except for one significantly massive disk in NGC1333IRAS4A (we omit asymmetric objects in the paper).
ALMA observations, on the other hand, reported smaller value of disk mass for Class 0/I YSOs.
The disk mass of L1527 \citep{2014ApJ...796..131O} and Lupus3 MMS \citep{2017ApJ...834..178Y}
are estimated as $1.3 \times 10^{-2} \msun$ and $1.0 \times 10^{-1} \msun$, respectively.
These estimated values (especially L1527 IRS) is smaller than the mass of gravitationally unstable disk.
Note, however,  that there are several uncertainties in the estimate of disk mass.
Dust opacity depends on dust size, composition, and
the shape \citep[e.g.,][]{1993Icar..106...20M,1994A&A...291..943O, 2018ApJ...869L..45B}.
Dust growth and subsequent dust 
radial drift possibly decreases the dust-to-gas mass ratio and 
cause underestimation of gas mass \citep{2017ApJ...838..151T}.
Dust scattering due to the dust growth may also decrease the dust thermal
emission, especially for the short wave length \citep{1993Icar..106...20M,2019ApJ...877L..18Z}.
Considering these uncertainties, we think that the formation of 
a massive disk in the early evolution phase
does not strongly contradict current observations.
However, longer term simulations and detailed comparisons with observations
are important subjects for future study.

\subsection{Condition of drift of magnetic field and ion in the envelope}
\label{discussion_in_drift}

  In \S \ref{B_drift}, we investigated
  magnetic field drift induced by ambipolar diffusion in the envelope.
  We showed that magnetic field drift more easily occurs with relatively large dust
  grains in which the ambipolar diffusion is strong in the envelope
  (see figure \ref{eta_fig}).
  We also confirmed that the Hall parameter is
  $\beta_{\rm Hall}> 1 $ in $r\gtrsim 100 $ AU in the simulations with magnetic field drift.
  The Hall parameter indicates the degree
  of ion-magnetic field coupling and $\beta_{\rm Hall} \gg 1$ means
  that the ion is well coupled to the magnetic field
  \citep[for details, see][]{2002ApJ...573..199N}.
  Thus, our results suggest that ion-neutral drift may occur in
  the envelope of Class 0/I YSOs, especially with relatively large dust grains.
  Figure \ref{rdrift} shows that the region with outward
  magnetic field drift has the size of $r_{\rm drift} \lesssim 100$ AU when
  $M_{\rm star}+M_{\rm disk} \lesssim 0.1 \msun$ and expands to $r_{\rm drift} > 100$ AU
  as the mass in the center increases.

Recently, \citet{2018A&A...615A..58Y} attempted to 
observe ion-neutral drift in young Class 0 YSO B335.
They did not detect ion-neutral drift (more precisely,
the velocity difference of ion and neutral is less than $\delta v<0.3 \kms$).
Possible explanations for this non-detection is that B335 is too young
($M_{\rm star} \sim 0.05\msun$) or that the dust size
in the envelope is not large enough.
We suggest that a more evolved Class 0/I YSOs with
a central protostar (plus disk) mass of $M>0.1 \msun$ may be a good
candidate to observe ion-neutral drift.
If a velocity difference of ion and neutral is observed in future observations,
it will provide a unique opportunity to quantify the magnetic field strength because
ion-neutral drift velocity is a function of magnetic field strength.



\subsection{Formation and early evolution of outflow}
In our simulations, outflows are ubiquitously formed.
At the end of the simulation, its size reaches
several $1000$ AU (figure \ref{outflow}).
We found that the mass, linear momentum, and angular momentum of outflow
depend on the dust model.
The outflows in small dust models
(model\_a0035$\mum$, model\_a01$\mum$, and model\_MRN) tend to have
larger mass, linear momentum, and angular momentum than those
in large dust models (model\_a03$\mum$ and model\_trMRN).
The difference of $\eta_A$ in the low-density outflow region may cause this difference.

The mass of outflow formed in our simulation is in the range 
of $10^{-2} \msun < M < 10^{-1} \msun$ (figure \ref{outflow}).
This is in good agreement with observations.
\citet{2004A&A...426..503W} reported that an observed outflow with dynamical
time of $\lesssim 10^4$ yr has 
a mass of  $10^{-3} \msun<M<10^{-1} \msun$ and mostly within
the range of $10^{-2} \msun<M<10^{-1} \msun$. 
The dispersion of the outflow mass reported in  \citet{2004A&A...426..503W}
is possibly due to the difference in 
ionization degree and hence difference of typical dust size.

\section{Summary}
Our results are summarized as follows.

\begin{enumerate}
\item
  Circumstellar disks are formed in all simulations.
  The disks have sizes of  $\sim 10$ AU at the protostar formation epoch
  and grow to several tens of AU 
  at $ \sim 10^4$ yr after protostar formation.
  Disk sizes are almost identical among the simulations as long as
  pseudo-disk warp does not develop.
  Once pseudo-disk warp develops, the disk begins to shrink.
\item
  Magnetic field drift in the envelope may occur in 
  the early evolution of young stellar objects.
  The Hall parameter in the envelope is generally $\beta_{\rm Hall} \gg 1$,
  and ion-neutral drift is also expected there.
  Ion-neutral field drift of 
  $\vel_{\rm drift} \gtrsim 0.19 \kms$ at $r\gtrsim 100$ AU 
  occurs under conditions with
  relatively large dust grains of $a \gtrsim 0.2 \mum$ (or absence of small grain) 
  and protostar (plus disk) mass of  $M \gtrsim 0.1 \msun$.
\item
  The mass of the circumstellar disk tends to be
  comparable to the mass of the central star,
  and gravitational instability develops in the early phase of disk evolution.
  A massive disk is a consequence of the high mass accretion rate at
  the early evolution stage.
\item
  The warp of the pseudodisk can develop at $\sim 10^4$ yr after protostar formation.
  The warp enhances magnetic field strength and magnetic braking in the disk, and
  has a negative impact on disk growth.
\item
  Outflows are ubiquitously formed.
  In some simulations, its angular momentum becomes comparable
  to the disk angular momentum, and outflow may have
  a major impact on disk growth.
  \end{enumerate}

\section*{Acknowledgments}
We thank Dr. Iwasaki Kazunari 
and Dr. Okuzumi Satoshi for fruitful discussions.
We also thank anonymous referee for helpful comments.
The computations were performed on a parallel computer, XC40/XC50 
system at CfCA of the NAOJ.
This work is supported by JSPS KAKENHI 
grant number 17H06360, 18H05437, 18K13581, 18K03703.

\appendix

\section{Initial density and magnetic field configuration}
\label{analytic_arg}
In this appendix, we describe our initial and boundary conditions
in detail, as well as the motivations for adopting the initial conditions.

In long-term simulations of cloud core collapse after protostar formation,
the outflows grow to the scale of $\gtrsim 10^3$ AU i.e.,
comparable to the initial radius of the core.
Thus, precise care is required for the outer boundary.
In our previous studies, we adopted a rigidly rotating shell at $r\sim R_c$,
where $R_c$ is the radius of the cloud core.
However, our numerical experiences has shown
that such a boundary reflects the outflow and shakes up the
density structures in $r< R_c$, which is clearly numerical.
Thus, we need more appropriate outer boundary condition.

Our strategy follows that of \citet{2011PASJ...63..555M}, i.e.,
setting the outer boundary far from the cloud core surface
by adding surrounding medium to the core  \citep[see e.g.,][for different strategy
to impose outer boundary with SPH]{2007Ap&SS.311...75P}.
\citet{2011PASJ...63..555M} placed the molecular cloud core in 
a medium with a constant density.
The size of medium was $2^5 R_c$.
With a nested grid code (or AMR code), the outer medium only 
requires acceptable computational costs.
However,  with the SPH scheme, the computational cost is proportional
to the mass, and hence volume for a constant density medium,
and a $2^5$ times larger mass requires unacceptably large computational 
costs.

To avoid this problem, we adopted the Bonner-Evert sphere surrounded by a
medium with a steep density profile of $\rho \propto r^{-4}$ for $r>R_c$ as
described in \S \ref{init_condition}.
With this profile the total mass of the entire domain is
 $\sim 2 M_c$, even when we set the boundary radius to $R_{\rm b}=10 R_c$.

A problem then arises for the magnetic field structure.
If we adopt a constant magnetic field with our density profile,
the plasma $\beta$ obeys $\beta\propto r^{-4}$ in the outer medium,
and a low $\beta$ region emerges, which requires very small time-stepping.
To avoid small time-stepping, we constructed a magnetic field profile
which has a constant vertical component in the central region,
and decreases in the outer region.
With this magnetic field profile, plasma $\beta$ becomes constant in the larger radius
and the low $\beta$ problem is avoided.

The magnetic field of our initial conditions is generated 
from the vector potential in cylindrical coordinate $(R,z)$ of 
\begin{eqnarray}
A_\phi=\frac{B_0 R_0}{2(\frac{R}{R_0}+\frac{R_0}{R}+\frac{z^2}{R R_0})}.
\end{eqnarray}
and the resultant magnetic field is 
\begin{eqnarray}
B_R&=&\frac{B_0}{(1+(R/R_0)^2+(z/R_0)^2)^2} \frac{R z}{R_0^2},\\
B_z&=&\frac{B_0 (1+(z/R_0)^2)}{(1+(R/R_0)^2+(z/R_0)^2)^2}.
\end{eqnarray}
where $B_0$ is the magnetic field at the center.

This magnetic field profile has the  desired nature.
In $R\to0$, the magnetic field becomes constant and has only the $z$ component.
In the spherical coordinate $(r,\theta)$, the magnetic field strength is given as
\begin{eqnarray}
  |\mathbf{B}|=\frac{B_0}{\sqrt{2}((\rR)^2+1)^2} \\
  \sqrt{2+2(\rR)^2+(\rR)^4 +(\rR)^2\left\{2+(\rR)^2\right\} \cos 2 \theta} \nonumber
\end{eqnarray}
and except at the midplane ($\theta=\pi/2$), 
$B \propto r^{-2}$ as $r \to \infty$.
On the other hand, $B \propto r^{-4}$ as $r \to \infty$ at the midplane.
The ratio of the $R$ and $z$ components of the magnetic field is given as 
\begin{eqnarray}
\frac{B_R}{B_z}=\frac{(\rR)^2 \cos \theta \sin \theta}{1+(\rR)^2 \cos^2 \theta},
\end{eqnarray}
and 
\begin{eqnarray}
\frac{B_R}{B_z}=\tan \theta=\frac{R}{z}~(r\to \infty).
\end{eqnarray}
Thus, the magnetic field is parallel to the position vector ($\magB \parallel \mathbf{r}$) as
$r\to \infty$ apart from at the midplane.
On the other hand, $B_R=0$ at the midplane and the magnetic field has 
only a vertical component at the midplane.
With our magnetic field configuration, 
the magnetic flux is 1/2 times smaller at the edge of the core $R=R_0$ than that 
with constant magnetic field with $B_0$ although the central magnetic field strength
is the same.

  \section{Numerical tests on the origin of the Warp}
  \label{appendix_warp}
  As we have shown, the warp of the pseudodisk often develops in our simulations.
  To confirm that the warp is not due to the numerical artifact but is physical,
  we conducted two numerical tests.
  In this appendix, we describe the results of the numerical tests and
  show that the warp develops even without a sink particle
  and even in a simulation with a nested-grid code.

  When we first obtained the warp,
  we were concerned that the numerical
  artifact of the sink particle possibly causes the warp.
  To deny this possibility,
  we conducted the simulation without the sink particle but
  employing stiff EOS (in other words, we keep using
  the equation (\ref{EOS}) which is not appropriate in high density of
  $\rho\gtrsim 10^{-11} \gcm$).

\begin{figure}
\includegraphics[clip,trim=0mm 0mm 0mm 0mm,width=45mm,,angle=-90]{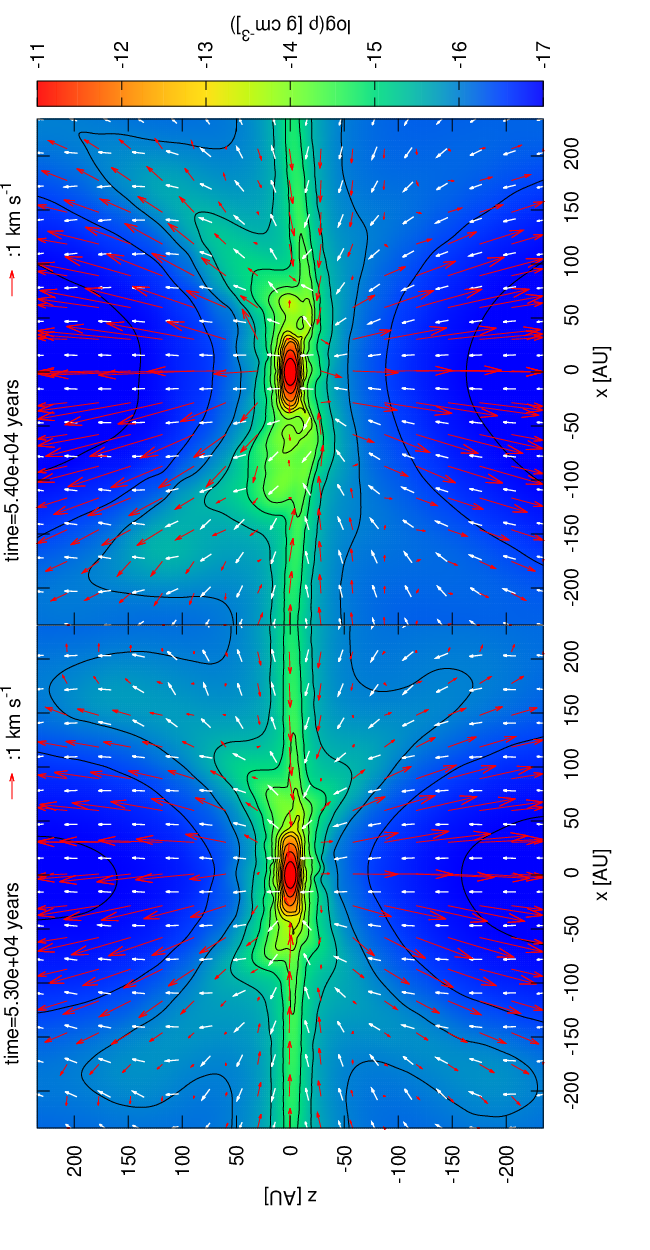}
\caption{
  Density cross-sections on the $x$-$z$ plane for central 500-AU
  square region of the simulations with the stiff EOS before and after the development
  of the warp.
  The initial conditions and dust model are the same as model\_MRN.
  }
\label{fig_stiff_EOS}
\end{figure}

  The density cross-section of the simulation is shown in figure \ref{fig_stiff_EOS}.
  The initial condition and the dust model are the same as the model\_MRN.
  The figure clearly shows that the warp also develops even without a sink particle,
  although we find that the epoch of warp formation is slightly ($\sim 10^3$ yr) delayed.
  Thus, we conclude that the warp is not caused by the numerical artifact of the sink particle.

Another concern was the possible artifact due to the numerical scheme.
Although the Godunov SPMHD scheme passes major numerical
tests very well \citep[see,][]{2011MNRAS.418.1668I}
and we are sure that it can reasonably capture the evolution of the cloud core,
the additional test with another numerical scheme was desired.

For this purpose, we conducted a disk formation simulation with the 
numerical code which used in \citet{2019ApJ...876..149M}.
The simulation settings of this simulation are different
from those of the other simulations presented in this paper.
We briefly describe the settings of the simulation below.
The simulation was done with the nested grid code which has been developed
by Machida and his collaborators. As the initial condition, the cloud core with a Bonnor-Ebert
density profile was adopted. The initial cloud core has a radius of $5.9 \times 10^3$ AU
and a mass of $1 \msun$.
A uniform magnetic field of $B_0 = 43~ \mu{\rm G} $ and
a rigid rotation of $\Omega_0 = 1.5 \times 10^{-15}~ {\rm s}^{-1}$
are added to the initial cloud core, which correspond to the normalized mass-to-flux ratio of $\mu=3$
and the ratio of rotational to gravitational energy of $ \beta_{\rm rot} = 0.02$.
As the cloud collapses, a finer grid is automatically generated
to ensure the Truelove condition, in which the Jeans wave
length is resolved at least 16 cells. Each rectangular grid has cells of (i, j, k) = (64, 64, 64).
The sink cell technique was used with a threshold density of $10^{14} {\rm cm^{-3}}$
and a sink accretion radius $0.5$ AU.
The grid size and cell width of the finest grid are $24$ AU and $0.36$ AU, respectively.
Equations (1)-(7) in \citet{2020MNRAS.491.2180M} were solved for this simulation.

\begin{figure}
\includegraphics[clip,trim=0mm 0mm 0mm 0mm,width=80mm]{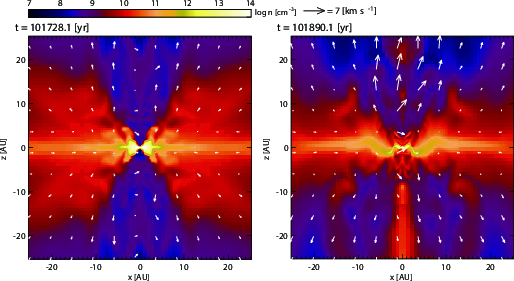}
\caption{
  Density cross-sections on the $x$-$z$ plane for central 50-AU
  square region of the simulations with a nested grid code before
  and after the development of the warp.
  }
\label{fig_grid}
\end{figure}

The figure \ref{fig_grid} shows the simulation results and the warp 
is also formed in this simulation although its size is small compared to
the other simulations in this paper, which may reflect the fact that
the magnetic field is stronger and the rotation is weaker than the other simulations
in this paper.
Despite we employed completely different numerical scheme and initial condition,
we reproduced the warp.
We believe that the reproduction of the warp
in these numerical tests strengthens
the claim that the warp has physical origin.

\bibliography{article}

\end{document}